\documentclass{aa}  
\usepackage{graphicx}
\usepackage{txfonts}
\usepackage{longtable}
\usepackage{lscape}

\usepackage{natbib,twoopt}
\usepackage{hyperref} 
\hypersetup{breaklinks=blue,colorlinks=blue}

\bibpunct{(}{)}{;}{a}{}{,} 
\makeatletter
\newcommandtwoopt{\citeads}[3][][]{\href{http://adsabs.harvard.edu/abs/#3}%
{\def\hyper@linkstart##1##2{}%
\let\hyper@linkend\@empty\citealp[#1][#2]{#3}}}
\newcommandtwoopt{\citepads}[3][][]{\href{http://adsabs.harvard.edu/abs/#3}%
{\def\hyper@linkstart##1##2{}%
\let\hyper@linkend\@empty\citep[#1][#2]{#3}}}
\newcommandtwoopt{\citetads}[3][][]{\href{http://adsabs.harvard.edu/abs/#3}%
{\def\hyper@linkstart##1##2{}%
\let\hyper@linkend\@empty\citet[#1][#2]{#3}}}
\newcommandtwoopt{\citeyearads}[3][][]%
{\href{http://adsabs.harvard.edu/abs/#3}
{\def\hyper@linkstart##1##2{}%
\let\hyper@linkend\@empty\citeyear[#1][#2]{#3}}}
\makeatother

\begin{document}

   \title{Precision Electron Measurements in the Solar Wind at 1~au from NASA's Wind Spacecraft}


   \author{Chadi S. Salem\inst{1}
          \and
          Marc Pulupa\inst{1}
          \and
          Stuart D. Bale\inst{1,2}
          \and
          Daniel Verscharen\inst{3,4}
          }

   \institute{Space Sciences Laboratory, University of California, Berkeley, CA 94720, USA\\
              \email{salem@ssl.berkeley.edu}
         \and
         Physics Department, University of California, Berkeley, CA 94720, USA
         \and
             Mullard Space Science Laboratory, University College London, Dorking, RH5~6NT, UK
        \and
            Space Science Center, University of New Hampshire, Durham, NH 03824, USA
             }

\titlerunning{Precison Electron Measurements in the Solar Wind at 1~au}

 
  \abstract
   {The non-equilibrium characteristics of electron velocity distribution functions (eVDFs) in the solar wind are key in understanding the overall plasma thermodynamics as well as the origin of the solar wind.  More generally, they are important in understanding heat conduction and energy transport in all weakly collisional plasmas. Solar wind electrons are not in Local Thermodynamic Equilibrium (LTE), and their multi-component eVDFs develop various non-thermal characteristics, such as velocity drifts in the proton frame, temperature anisotropies as well as suprathermal tails and heat fluxes along the local magnetic field direction.}
   {This work aims to characterize precisely and systematically the non-thermal characteristics of the eVDF in the solar wind at 1\,au using data from the Wind spacecraft.}
   {We present a comprehensive statistical analysis of solar wind electrons at 1\,au using the electron analyzers of the 3D-Plasma instrument on board Wind. This work uses a sophisticated algorithm developed to analyze and characterize separately the three populations -- core, halo and strahl -- of the eVDF up to {\it super-halo} energies (2 keV). This algorithm calibrates these electron measurements with independent electron parameters obtained from the  quasi-thermal noise around the electron plasma frequency measured by Wind's Thermal Noise Receiver (TNR).
   The code determines the respective set of total electron, core, halo and strahl parameters through non-linear least-square fits to the measured eVDF, taking properly into account spacecraft charging and other instrumental effects, such as the incomplete sampling of the eVDF by particle detectors.}
   {We use four years, approximately 280\,000 independent measurements, of core, halo and strahl electron parameters to investigate the statistical properties of these different populations in the slow and fast solar wind. We discuss the distributions of their respective densities, drift velocities, temperature, and temperature anisotropies as functions of solar wind speed. We also show distributions with solar wind speed of the total density, temperature, temperature anisotropy and heat flux of the total eVDF, as well as those of the proton temperature, proton-to-electron temperature ratio, proton-$\beta$ and electron-$\beta$. Intercorrelations between some of these parameters are also discussed.}
   {The present dataset represents the largest, high-precision, collection of electron measurements in the pristine solar wind at 1~AU. It provides a new wealth of information on electron microphysics. Its large volume will enable future statistical studies of parameter combinations and their dependencies under different plasma conditions. }


   \maketitle
%

\section{Introduction}

The contribution of electrons to the generation, acceleration and evolution of the solar wind is still a critical and unsolved problem in solar wind physics \citep{Marsch:2006, Verscharen:2019b}. For the development of predictive, physics-based solar wind models, a more accurate understanding of electron physics is crucial \citep{Scudder:2019c}.  

Electrons are the most abundant particle species in all fully-ionized plasmas like the solar wind. Despite their negligible contribution to the total solar wind momentum flux, they have a significant impact on the overall dynamics and thermodynamics of the solar wind for multiple reasons: (1) their high mobility adjusts quasi-neutrality on very short time scales compared to any other plasma time scales \citep{Feldman:1975}; (2) their pressure gradient creates a significant electric field, which can lead to ambipolar diffusion and electrostatic acceleration \citep{Lemaire:1971, Lemaire:1973, Scudder:1992, Scudder:1994, Scudder:1996, Maksimovic:1997, Maksimovic:2001, Zouganelis:2004, Zouganelis:2005, Pierrard:2012, Scudder:2019b}; and (3) the observed non-zero skewness of the electron velocity distribution function (eVDF) provides the solar wind with a significant heat flux \citep[e.g.][]{Feldman:1976, Scime:1994b, Scime:1999, Scime:2001, Salem:2003, Bale:2013, Halekas:2020, Halekas:2021}.
 
The kinetic properties of the electrons are fully described by the eVDF.  Indeed, solar wind eVDFs have a complex structure that  evolves with heliocentric distance. This  evolution is the result of a delicate balance between several competing processes involving the Sun's gravitation, the magnetic field, the electric field, Coulomb collisions as well as wave--particle interactions \citep[with background electromagnetic turbulence and/or via plasma micro-instabilities;][]{Verscharen:2019b}. This balance is restated by the Generalized Ohm's Law (GOL), a steady-state electron equation of motion in the ion frame of reference  \citep{Rossi:1970, Scudder:2019a, Scudder:2019b}.  Information on these various processes is embedded in the eVDFs, with direct implications for electron heat conduction, as well as the energy budget and the physics of the expanding solar wind \citep{Hollweg:1974, Hollweg:1976, Cranmer:2009, Stverak:2015, Scudder:2019b}.

\begin{figure}[ht]
  \centering
   \includegraphics[width=0.8\hsize]{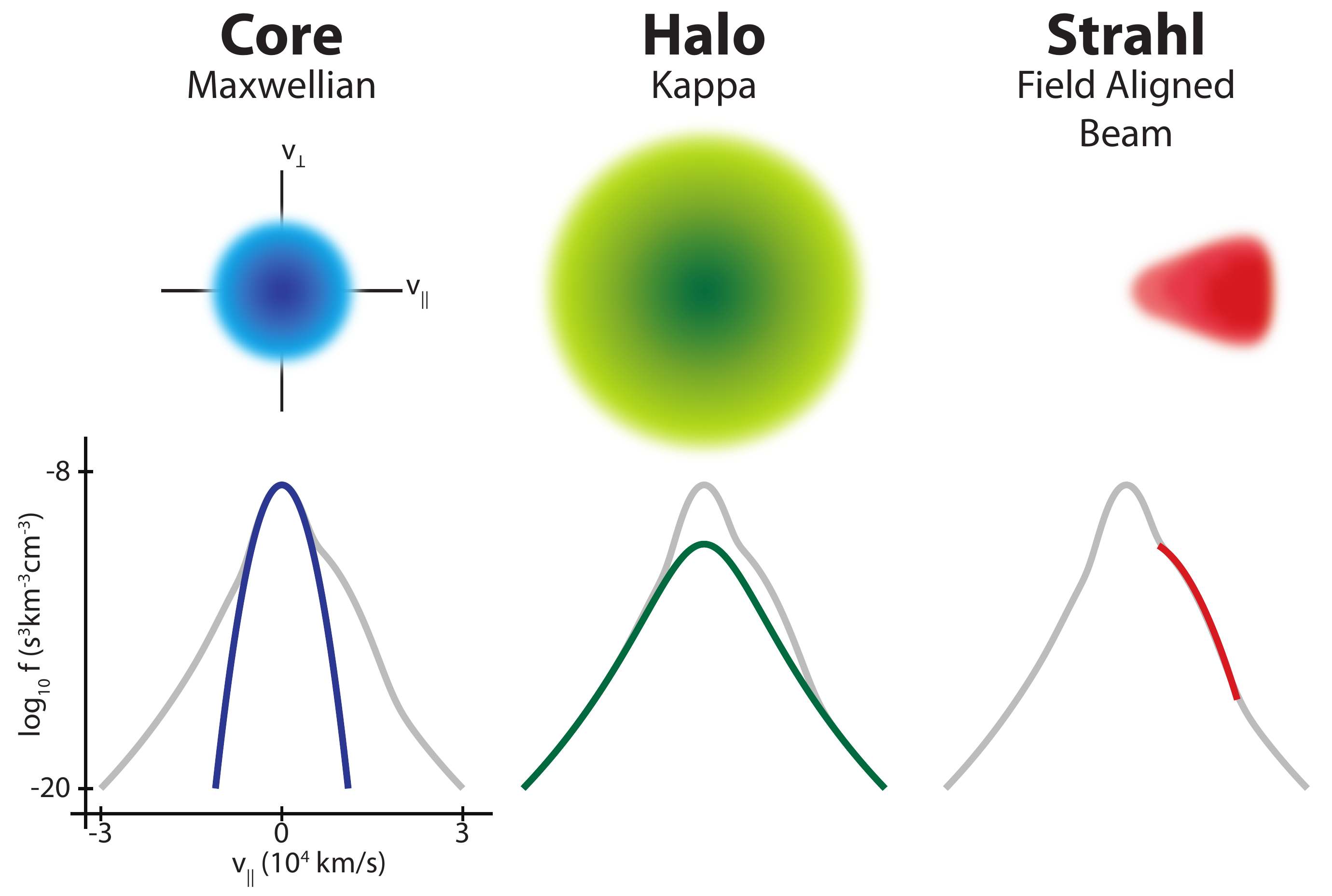}
\caption{\small Illustration of the three solar wind electron populations: the core, halo and strahl. The top row shows a 2-dimensional view of the eVDFs of these three components, and the bottom row highlights each component in parallel cuts through the eVDFs.}
\label{fig_cartoon}
\end{figure}

Observations show that the solar wind eVDF at 1\,au consists of three main components \citep{Feldman:1975, Rosenbauer:1977, Pilipp:1987, Pilipp:1987c, Hammond:1996, Lin:1998, Fitzenreiter:1998, Maksimovic:2000, Gosling:2001, Salem:2003, Maksimovic:2005,Salem:2007,Stverak:2009,Halekas:2020}: a primary cool thermal {\it core} ($\sim 10$\,eV, $\sim 95\%$ relative density), a superthermal {\it halo} ($\sim 50$\,eV, $\sim 4\%$ relative density), and a field-aligned anti-Sunward beam called {\it strahl} ($\sim 100-1000$\,eV, $\sim 1\%$ relative density). This core-halo-strahl structure is illustrated in Figure \ref{fig_cartoon}. The top row shows a 2D view of the eVDFs of these three components in the ion frame, projected in a plane parallel and perpendicular to the local magnetic field $[v_{\parallel},v_{\perp}]$. The bottom row highlights each component in a parallel cut through their eVDFs. The core is well described by a bi-Maxwellian distribution, while the halo population is well described by a (bi-)$\kappa$-distribution with large velocity tails in the eVDF \citep{Feldman:1975,Maksimovic:2005,Stverak:2009,Wilson:2019a,Wilson:2019b}. The strahl has a more complicated cone-shaped structure, with an angular width that is highly variable between slow and fast solar wind, evolves with distance from the Sun, and depends on energy \citep{Pilipp:1987, Pilipp:1987c, Hammond:1996, Ogilvie:1999, Anderson:2012, Gurgiolo:2016, Graham:2017, Horaites:2018a, Bercic:2019, Bercic:2020, Halekas:2020}.

Each population often exhibits  temperature anisotropies and rather large relative drifts parallel or anti-parallel to the local magnetic field with respect to the other populations and the solar wind ions \citep{Feldman:1975, Stverak:2008, Pulupa:2014}.  The three populations and their temperature anisotropies are more distinctive in the fast solar wind. This observation is usually attributed to collisional effects, which are stronger in the slow solar wind \citep{Salem:2003, Marsch:2006}. Although the individual drift speeds of the core, halo and strahl relative to the ions satisfy globally quasi-neutrality and the zero-current condition, they correspond to a substantial heat flux, usually directed anti-Sunward.
On a larger scale, studies of the radial evolution of eVDFs show that the relative density of the halo increases radially while that of the strahl decreases at heliocentric distances greater than 0.3\,au \citep{Maksimovic:2005, Stverak:2009, Halekas:2020}. At the same time, the angular width of the strahl increases with distance \citep{Hammond:1996, Graham:2017, Bercic:2019, Bercic:2020}.  
These results suggest that the halo electrons are strahl electrons that have been backscattered and pitch-angle diffused by mechanisms that remain unidentified. 

Both collisional and/or collisionless processes have been suggested to be responsible for controlling these electron non-thermal properties, via Coulomb collisions \citep{Scudder:1979, Scudder:1979b, Phillips:1990, Lie:1997, Landi:2003, Salem:2003, Stverak:2008, Bale:2013, Landi:2012, Landi:2014, Horaites:2015, Horaites:2019, Boldyrev:2019, Bercic:2021} and/or (resonant or non-resonant) wave--particle interactions  \citep{Gary:1975, Gary:1994, Gary:1999, Gary:1993, Krafft:2003, Vocks:2005, Gary:2007, Shevchenko:2010, Seough:2015, Kajdic:2016, Horaites:2018b, Roberg-Clark:2018a, Roberg-Clark:2018b, Roberg-Clark:2019, Verscharen:2019, Vasko:2019, Jeong:2020, Micera:2020, Jagarlamudi:2021, Innocenti:2020, Cattell:2021} respectively. On the other hand, a more recent theoretical paradigm \citep{Scudder:2019b} predicts that the non-thermal shape of the eVDFs is a corollary of the strong parallel electric field needed to enforce the GOL balance so that no currents flow in the plasma, in a variant of Dreicer's transient runaway process \citep{Dreicer:1959,Dreicer:1960, Fuchs:1986}.

At energies above those of the halo and strahl, the eVDF at times exhibits a {\it super-halo} \citep{Lin:1980, Wang:2012, Wang:2015}, a fourth population with energies from about 2 to 200\,keV. The super-halo seems to be a quasi-isotropic and steady-state feature of the solar wind, although its origin still remains unknown \citep{Yang:2015}. 

To date, our understanding of the processes that regulate the solar wind electron properties and how the electrons couple to the ions are not fully understood \citep[e.g.][]{Marsch:2006, Verscharen:2019b}.  This work aims to {\it accurately} and {\it systematically} characterize and quantify the non-thermal features of solar wind eVDFs and extract the properties of their core, halo and strahl components using data from NASA's Wind spacecraft (s/c) \citep{Acuna:1995, Harten:1995}. Wind is best-suited for this study, with its well-designed, high resolution, particle and field instrumentation, the 3D Plasma \citep{Lin:1995} and the WAVES \citep{Bougeret:1995} experiments respectively, and over 26 years of continuous data collection (i.e., over 2.5 solar cycles worth of data) to allow statistically significant studies of the variation and evolution of the solar wind electron properties.
This is key to shed light on the underlying physical processes at play to control the non-thermal eVDF shapes and properties, and ultimately solve the electron physics puzzle in the solar wind and in electron astrophysics in general \citep{Verscharen:2021}.

Precision measurements of electrons are crucial for understanding the thermodynamics and microphysics of the plasma. These measurements are difficult to make and interpret: the low mass of electrons renders them especially susceptible to s/c charging effects in addition to inherent limitations in instrument capabilities. Indeed, the eVDF in the vicinity of the s/c is severely polluted at low energies (a few eV) by photo-electrons and secondary electrons emitted by the s/c body, and distorted by the electric field created by s/c charging. This electric field modifies both the energy and the direction of motion  incident solar wind electrons  \citep{Grard:1973, Garrett:1981, Whipple:1981, Goertz:1989, Scime:1994, Pulupa:2014}. Photo-electrons and secondary electrons contribute artificial counts to measured electron spectra at energies lower or equal to the energy associated with the  s/c electric potential $\phi$. Correcting for these effects is a difficult task since $\phi$ varies with the electron properties (density, temperature, etc.) that one wishes to measure. These corrections are complicated by the fact that the eVDF is measured within a finite energy range [$E_{\min},E_{\max}$] \citep{Song:1997, Salem:2001, Genot:2004}. For typical electron detectors in the solar wind,  $E_{\min} \sim 5$ to $10$\,eV, usually above the varying $e\phi$, so a variable part of the electron core is missing \citep{Salem:2001}. 

There are various ways of addressing these issues, the most natural being proper instrumentation measuring the s/c electric potential itself. Direct on-board measurements of the s/c potential using DC electric field instruments \citep{Cully:2007} have only been implemented on missions such as Cluster, THEMIS, MMS, Parker Solar Probe and Solar Orbiter. Older space missions such as Helios, Wind or Ulysses do not have such DC electric field measurements and analyzing their electron data requires very careful handling and assessment of s/c charging and other instrumental effects. 

\citet{Salem:2001} discuss the various ways to tackle these effects in the absence of direct s/c potential measurements and propose an alternative, robust, technique to calibrate data from the electrostatic analyzers. These methods use data from ultra-sensitive radio receivers that measure the plasma thermal noise around the electron plasma frequency. From the plasma thermal noise, one can unambiguously derive  the total electron density \citep{MeyerVernet:1989, Issautier:1998, MeyerVernet:1998, Salem:2001, MeyerVernet:2017} as a reference. 

In this paper, we introduce an automated analysis method that corrects for the spacecraft potential using thermal-noise measurements from Wind's electric-field antennas, based \citet{Salem:2001}'s technique. We then fit the three component structure of the corrected electron spectra to analytical expressions for the eVDFs. Applying our method to 4 years of Wind data, we present a large statistical survey of 280\,000 reliable  eVDF parameters in the solar wind at 1\,au. This method and the resulting dataset will be important cornerstones for future studies of specific electron processes in the solar wind.

In Section 2, we describe the instrumentation and the data products that we use in this work. In Section 3, we present the different stages of the eVDF analysis technique. In Section 4, we present the results of our fit analysis, showing first the statistics of the total electron parameters, such as total density, temperature, temperature anisotropy and heat flux of the total eVDF, as functions of the solar wind speed. Section 4 also presents the statistical distributions of the three electron components -- core, halo and strahl -- separately, and intercorrelations between core, halo and strahl properties are also shown and discussed. Section 5 presents a discussion of our results on electron properties at 1\,au. Finally, Section 6 presents a conclusion synthesizing the new results from this work.

\section{Wind Instrumentation and Datasets}

 To perform the present work, we use electron and wave data, as well as solar wind plasma moments and magnetic field vectors from various instruments on the Wind s/c. Here we describe these instruments and associated data products.

Wind was launched in November 1994. With its complete package of high resolution wave and particle instruments and extended intervals near or at the Lagrange point L1, it has been an ideal and unique platform providing continuously  high time resolution -- wave and particle -- data ever in the solar wind at 1~AU.  Wind has now provided more than two solar cycles worth of data, allowing for comprehensive studies of waves, wave--particle coupling and of the microphysics of the solar wind in general \citep{Wilson:2021}. We consider data from free solar wind intervals (see next section) only, so the data presented here concerns the physics of the pristine, unperturbed solar wind.

We use full 3-D electron distribution functions from the Electron ElectroStatic Analyzers (EESA), part of the 3D-Plasma (3DP) experiment \citep{Lin:1995}. We also use plasma wave spectra around the electron plasma frequency from the Thermal Noise Receiver (TNR), part of the WAVES experiment \citep{Bougeret:1995}. These are used for the Quasi-Thermal Noise (QTN) spectroscopy technique (described below) to determine  electron parameters independently of the 3DP spectra in order to calibrate the 3DP electron data and estimate the s/c floating potential \citep{Salem:2001}.

In addition, we use solar wind plasma moments, namely solar wind speed and proton temperature, from the 3DP ion analyzers \citep{Lin:1995}, and the SWE Faraday Cup \citep{Ogilvie:1995}, as well as spin-resolution (3\,s) magnetic field data from the Magnetic Field Investigation (MFI) \citep{Lepping:1995}.


%
\subsection{The EESA electrostatic analyzers}

The pair of EESAs make three-dimensional measurements of eVDFs from 3\,eV to 30\,keV with a high time resolution, high sensitivity, wide dynamic range and a good energy and angular resolution. The EESA-Low (EESA-L) analyzer covers the range of 3\,eV to 1.1\,keV, with a smaller geometric factor than the EESA-High (EESA-H) analyzer, which covers the energy range 300\,eV to 30\,keV. Both instruments have operational fields of view of 
$180^\circ$ and 15 logarithmically spaced energy channels. They sweep out $4\pi$ steradians in one spacecraft spin (spin period of 3\,s). The data are combined on board into 88 angular bins for both instruments. Because of the available telemetry rate, the full 3D eVDFs are transmitted only every $\sim$96\,s when Wind is at L1, but onboard computed moments (calculated from the EESA-L distributions, up to the 3rd order) are transmitted with a cadence of 3\,s.

  The s/c charging effects generally combine with other instrumental effects such as the incomplete sampling of the VDFs due to a nonzero low-energy threshold of the energy sweeping in the electron 
spectrometer \citep{Song:1997, Salem:2001}.  Consequently, all the moments of 
the eVDFs are affected as well \citep[][and references therein]{Salem:2001}. 
A careful analysis of solar wind eVDFs requires therefore a full correction of these effects.


\subsection{The Thermal Noise Receiver}

The TNR is a very sensitive digital spectrum analyzer designed  to measure the electron QTN around the electron plasma frequency. The QTN arises from the thermal motion of the ambient ions and electrons, which produces electrostatic fluctuations that can be measured with a wave receiver connected to, for example, a wire dipole antenna. On Wind, the TNR is connected to the $2\times 50$\,m thin wire electric dipole antennas in the s/c spin plane, and consists of several separate spectral receivers that measure and digitize into five different (overlapping) frequency bands (A, B, C, D and E) of relative bandwidth ${\Delta}f/f = 4.3\%$) from 4 to 256\,kHz. Spectra are acquired every 4.5\,s in normal mode of operation (ACE mode).

\begin{figure}[ht]
  \centering
   \includegraphics[width=0.8\hsize]{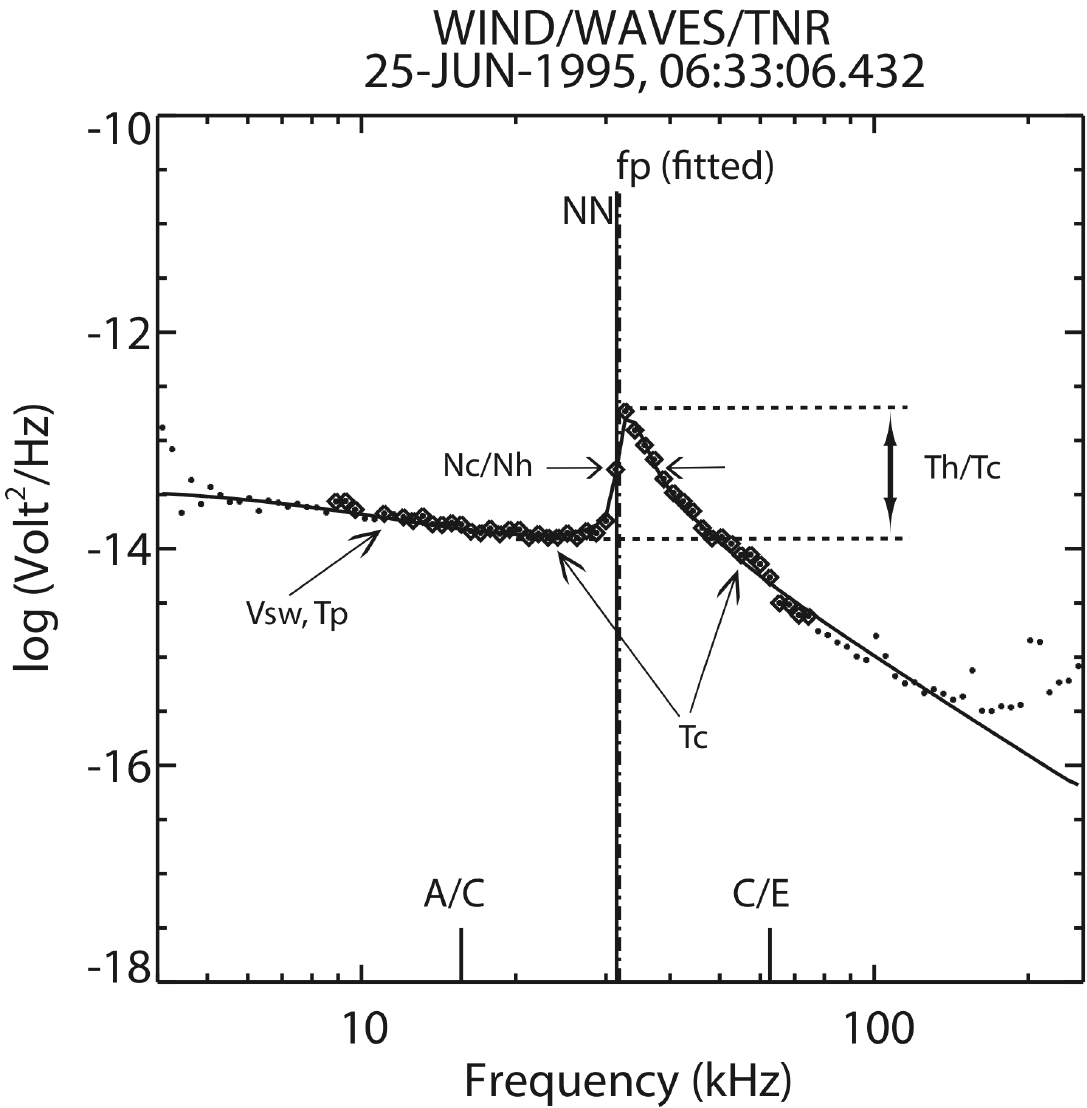}
\caption{\small Example of a typical voltage power spectrum around the electron plasma frequency measured by the Wind/WAVES/TNR instrument in the solar wind on June 25, 1995.  The solid line is the predicted spectrum fitted to the selected data (in diamonds) among data not retained (dots) for the QTN fitting (see text). The annotations indicate the electron parameters obtained by fitting the theoretical spectrum  from the ion and the electron QTN, using a sum of two isotropic Mxwellians -- a core `c' and a halo `h' -- to the observed spectrum.  The vertical line indicate the locations of the local electron plasma frequency obtained by both a Neural Network (`NN', solid line) and by QTN fit (`fitted', dashed line).}
\label{fig_tnrfit}
\end{figure}

The electric field spectrum around the electron plasma frequency contains a wealth of infomation about the solar wind electron populations. Figure \ref{fig_tnrfit} shows a typical spectrum measured by the TNR in the fast solar wind on 1995-06-25. The measured spectrum -- power in Volts$^2$/Hz versus frequency in kHz -- is represented by the small plus signs. 
The peak above the local electron plasma frequency ($f_{pe}$, indicated by the vertical lines) is completely determined by the eVDF and the antenna characteristics \citep{MeyerVernet:1989, Issautier:1998, MeyerVernet:2017}. 
Resolving well this plasma peak requires an electric antenna (of length $L$) much longer than the Debye length $\lambda_D$, $L \gg \lambda_D$.   QTN measurements are largely immune to spacecraft potential due to the large volume sensed by the antenna compared to the volume affected by the s/c. 

The determination of the electron density, $N_{e,TNR} \propto f^2_{pe}$, relies basically on the identification of the ``plasma line" at $f_{pe}$ in the QTN spectrum. This is done routinely using a Neural Network (NN) developed by the WAVES team at the Paris Observatory \citep{Richaume:1996,Salem:2000}. The plasma frequency is determined with an accuracy of half a channel, which corresponds to a density accuracy of about $4.4\%$, independently of any calibrations or any model assumptions for the eVDF.

A more refined but model-dependent way to determine the electron density and other electron parameters is the QTN spectroscopy technique \citep{MeyerVernet:1979, MeyerVernet:1989, MeyerVernet:1998, Issautier:1999, MeyerVernet:2017}.  This method consists of predicting the voltage spectrum measured at the tips of the wire antennas at frequencies much above the local ion characteristic frequencies and the electron gyrofrequency using a model for the eVDF, usually a sum of two isotropic Maxwellians, a core (density $N_c$ and temperature $T_c$) and a halo (density $N_h$ and temperature $T_h$). The total predicted spectrum around $f_{pe}$ is the sum of the Doppler-shifted proton spectrum (depending on the solar wind proton  speed $V_{sw}$ and temperature $T_p$) and the electron spectrum \citep{Issautier:1998}. Then we fit the measured spectrum to the predicted spectrum to derive the following parameters: total density $N_e = N_c + N_h$, core temperature $T_c$, halo-to-core density ratio $\alpha = N_h/N_c$ and halo-to-core temperature ratio $\tau = T_h/T_c$.  This QTN technique has been implemented on Wind to routinely determine the QTN electron parameters \citep{Salem:2000, Salem:2001, Salem:2003b}, using $V_{sw}$ and $T_p$ measured by the 3DP experiment at 3~s cadence.

Figure \ref{fig_tnrfit} shows the QTN fit of the measured spectrum.  The diamonds are the measured data points selected for the fit.  The NN $f_{pe}$ and the one from the QTN fit agree very well (given by the inflection point in the spectrum, below the plasma line). The value of $T_c$  determines both the plateau below the plasma line and the slope of the spectrum above it. The width and amplitude of the plasma line represent a measure of the density and temperature ratios, $\alpha$ and $\tau$, respectively. For the spectrum shown in Fig. \ref{fig_tnrfit}, $V_{sw} = 761$\,km/s and $T_p = 47.5$\,eV. The QTN fit parameters are:  error on the total fit $\sigma = 2.4\%$, $f_{pe} = 22.24 (\pm 0.78\%)$,kHz so that $N_e = 6.13 (\pm 1.56\%)$\,cm$^{-3}$, $T_c \simeq 11.53 (\pm 10.37\%)$\,eV, $\tau \simeq 5.57 (\pm 15.78\%)$ and $\alpha = 0.059 (\pm 67.33\%)$. As shown in the above example, the QTN technique is robust in yielding an accurate determination of the electron density within $2\%$ and the core temperature within $10\%$; however, the uncertainties on the suprathermal (halo) density and temperature are much higher due to our underlying assumption of an isotropic Maxwellian in the  QTN calculations, which does not account for the halo suprathermal tails or the strahl \citep{Issautier:1998, MeyerVernet:1998, MeyerVernet:2017}. 

Both the Neural Network and the QTN fit have routinely been applied to the TNR data in order to obtain electron parameters. So far, NN TNR densities are available from the beginning of the mission in late 1994 until 2020.  Electron parameters from the QTN fit are only available from late 1994 to late 2004. These data sets are key to the calibration process of the 3DP eVDF data and to getting a good estimate of Wind's s/c potential. 
The QTN technique, by its robustness, has been applied on Ulysses \citep{Maksimovic:1995, Issautier:1998, Issautier:2001}, and more recently on BepiColombo \citep{Moncuquet:2006}, Parker Solar Probe \citep{Maksimovic:2020, Moncuquet:2020}, and Solar Orbiter \citep{Maksimovic:2005b,Maksimovic:2020b}.

\subsection{The SWE Faraday Cup and 3DP ion data}

All of the ion (proton and Helium) data from the Wind SWE Faraday Cups have now been analyzed and processed using a sophisticated and adaptive, nonlinear code developed by \citet{Maruca:2012}.  This work  has enabled revolutionary studies on the temperature anisotropy instabilities of protons \citep{Kasper:2002, Maruca:2011}, and of $\alpha$-particles \citep{Maruca:2012b}, as well as ion heating/thermalization \citep{Kasper:2008, Kasper:2013, Maruca:2013b}.  

We complement this SWE proton and Helium dataset from Wind/SWE with proton data from the proton sensor, PESAL, of the 3DP experiment \citep{Lin:1995}. The 3DP proton dataset includes continuous 3\,s proton moments -- density, velocity and temperature.


\section{Analysis Method of the eVDF}
\label{evdf_processing}

\subsection{Data selection criteria}

The starting point of our analysis is to select solar wind intervals, outside the Earth's magnetosphere and away from the bow shock and the ion/electron foreshock regions.
We use a standard model of the bow shock \citep{Slavin:1981} to determine when the s/c is outside the shock region and test for magnetic field connectivity to ensure the s/c is outside the foreshock.

Once we have identified undisturbed solar wind intervals throughout the Wind mission, we select all eVDFs available from both EESA-L and EESA-H. We determine one-count levels and remove measured background count rates for both analyzers. For EESA-H, we also remove angular bins that are polluted by solar UV, which tend to be bins in the Sunward direction. 

We select minimum and maximum energy channels for both analyzers. For EESA-L, a minimum of 5-10\,eV is considered depending on the detector's configured minimum energy channel $E_{\min}$ in order to remove the measured photo-electrons. 
For EESA-H a maximum energy channel is set at 2-3\,keV. Above these energies, high-energy super halo electrons dominate the eVDFs \citep{Lin:1998,Wang:2012}. The super halo electrons are not considered in this analysis.

\subsection{Spacecraft potential}\label{scpot}

The EESA-L and EESA-H distributions are converted from counts to phase space density using the instrument geometric factor and integration time. 
The next step is to correct for the effects of s/c potential on the measured eVDFs. 
We make an initial estimate of the s/c potential based on the measured density from the QTN, using a current balance model which accounts for photo-electrons, solar wind electrons and ions including the effects of their thermal motions and bulk flows \citep{Salem:2000,Salem:2001}.  The first estimate $\phi_0$ of the s/c potential uses a photo-electron temperature $T_{ph}$ of 2\,eV \citep[see Appendix in][]{Salem:2001}.

We then use $\phi_0$ to correct the energies of the eVDFs: $E^{'} = E - e\phi_0$. These new energies are then converted into velocities in 3D velocity space, after which 
we transform them to the solar wind frame in a field aligned coordinate system $[v_{\parallel},v_{\perp1},v_{\perp2}]$.  The solar wind frame is defined as the frame of zero ion current, ${\bf J}_i = N_i\,{\bf V_{sw}} = N_p\,{\bf V}_p + 2\,N_{\alpha}\, {\bf V}_{\alpha}$, where $p$ and $\alpha$ denote protons and $\alpha$-particles respectively. We neglect any minor ion contributions.

Our s/c potential correction is isotropic (monopole potential). \citet{Pulupa:2014} consider the case of non-isotropic potentials for the same dataset used here and show that the correction is small affects only odd moments.
For typical solar wind parameters, the correction is $\sim$2\% in the velocity and $\sim$4\% in the heat flux.


\subsection{Gridding and combining EESA-L and EESA-H eVDFs}


We grid the data structure for both EESA-L and EESA-H eVDFs using a Delaunay triangulation method to interpolate a two-dimensional eVDF onto a regularly spaced grid.  For this, we symetrize the eVDF with respect to $V_{\parallel}=0$ under the assumption of gyrotropy and no drifts in the direction perpendicular to the background magnetic field.  

We then combine EESA-L and EESA-H structures using the one-count levels for each to cut the eVDF for best overlap. For energies where EESA-L counts are 10 times the one-count level or greater, EESA-L data is used. For higher energies, the more sensitive EESA-H detector is used.  For a typical eVDF, this transition energy between EESA-L and EESA-H data corresponds to a velocity between $1\times 10^4$ and $2\times 10^{4}$\,km/s. We test the consistency between EESA-L and EESA-H over their common energy range.

The result is one full eVDF from a few eV to 2-3\,keV, sampling core, halo and strahl electrons. 
\begin{figure}[ht]
\centering
\includegraphics[width=0.8\hsize]{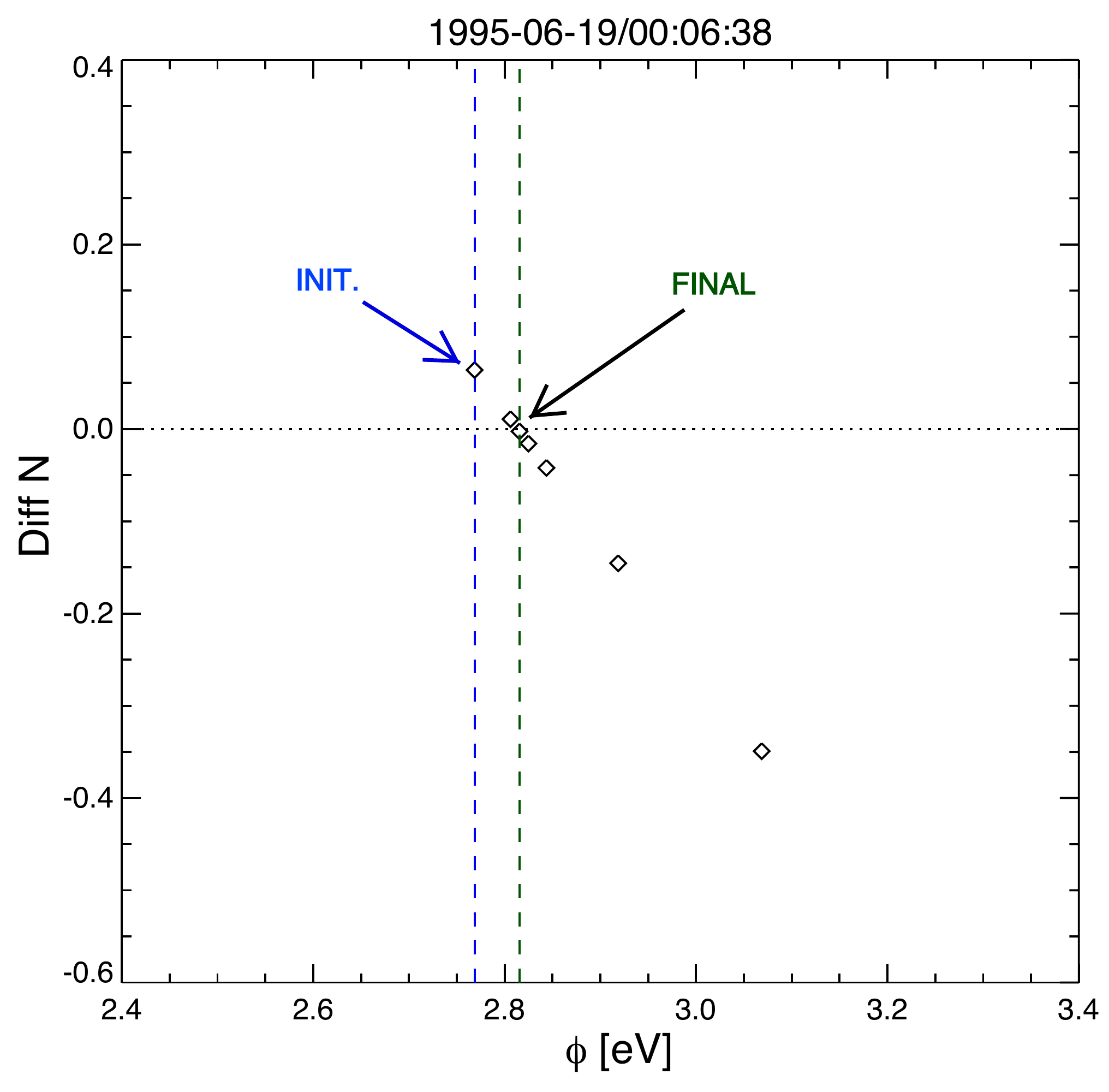}
\caption{\small Illustration of the iterative method for determining spacecraft potential (see text for details). We arrive at an estimated value $\phi$ of the s/c potential by iterating through various ``candidate'' s/c potentials, minimizing the difference between the electron density determined as the first-order moment of the eVDF and the electron density obtained from the QTN method.}
\label{fig_scpot}
\end{figure}

If the approximate s/c potential described in Section~\ref{scpot} is not the true s/c potential, this will introduce an error in the density moment of the eVDF \citep{Salem:2001}.  We do a first fit of the combined eVDF and calculate a total density, $N_{e3DP}$. Then, we test the accuracy of the s/c potential by comparing this total density to the highly accurate total density from the QTN method, $N_{eTNR}$.
If both densities are not equal, then we initiate an iterative s/c potential estimate and re-fit the eVDF until both 3DP and TNR density are equal (within a pre-defined margin), $N_{e3DP} = N_{eTNR}$  \citep{Pulupa:2014}. 
Figure \ref{fig_scpot} illustrates this iterative procedure to determine the true Wind s/c potential $\phi$ that we use in the final fit process to determine the  electron core, halo and strahl parameters. This true Wind s/c potential is retained as the one for which $N_{e3DP} = N_{eTNR}$.

 \subsection{eVDF fit technique}

The eVDF fit procedure consists of fitting cuts through the full eVDFs to a model of the solar wind electron distribution that includes two populations, the core and the halo, only. All fits are performed in IDL using a Levenberg--Marquardt least-squares fit technique \citep{2009ASPC..411..251M}.

The core is modeled by a drifting bi-Maxwellian distribution $f_c$, characterized by a density $n_{\mathrm{c}}$, parallel and perpendicular drift velocities $v_{\mathrm{c}\parallel}$ and $v_{\mathrm{c}\perp}$, and parallel and perpendicular temperatures $T_{\mathrm{c}\parallel}$ and $T_{\mathrm{c}\perp}$.  
The halo is modeled by a drifting bi-$\kappa$ distribution $f_h$, characterized by a density $n_{\mathrm{h}}$, parallel and perpendicular drift velocities $v_{\mathrm{h}\parallel}$ and $v_{\mathrm{h}\perp}$, parallel and perpendicular temperatures $T_{\mathrm{h}\parallel}$ and $T_{\mathrm{h}\perp}$, and a $\kappa$-parameter characterizing the high-energy power-law tails of the suprathermal halo \citep{Maksimovic:1997,Stverak:2009}.  The beam-like strahl population $f_s$ is not fitted to a closed  function; it is instead defined as the excess in the 2-D distribution function between the measured distribution $f$ and the core+halo model, i.e. $f_s = f - f_c - f_h$. 

\begin{figure*}[ht]
\centering
 \includegraphics[width=0.9\hsize]{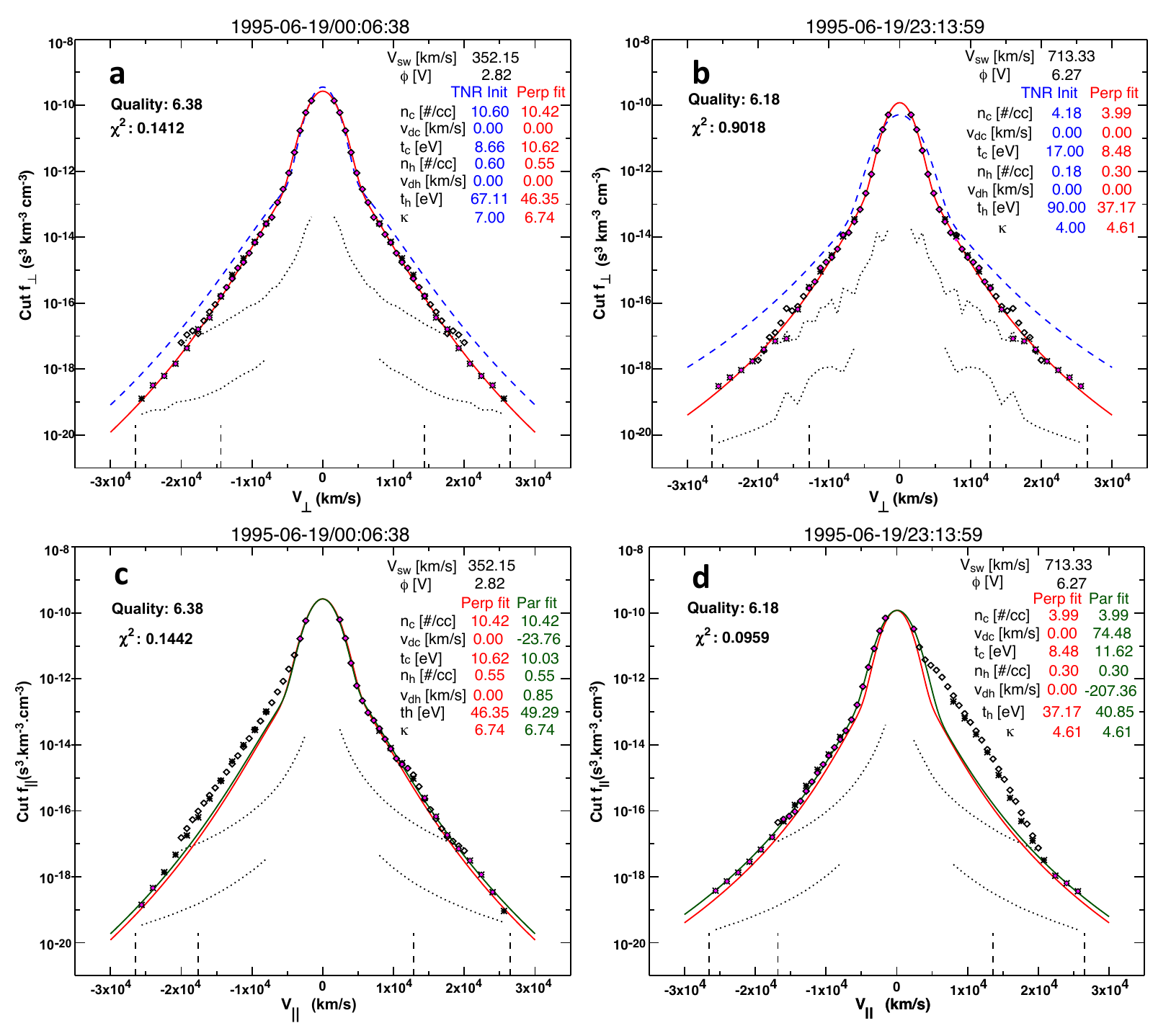}
\caption{\small Two typical eVDFs measured by EESA-L and EESA-H at 1\,au in the slow solar wind -- panels (a) and (c)
--, and in the fast solar wind -- panels (b) and (d). The top panels (a) and (b) show cuts through the eVDF in one of the two directions perpendicular to the local magnetic field {\bf B}: the diamonds are data points from EESA-L and the asterisks from EESA-H. The dotted lines represent the one-count levels for EESA-L and EESA-H. The blue dashed line represents the sum of Maxwellian and $\kappa$-distributions calculated using the QTN fit parameters (indicated in blue), which are used to initialize the  eVDF fit. The red line represents the fit to the measured perpendicular eVDF cut; the resulting fit parameters are indicated in red. The bottom panels (c) and (d) show cuts through the eVDF in the direction parallel to {\bf B}. The perpendicular fit is reported in red, and the perpendicular fit parameters are used to initialize the parallel eVDF fit, the results of which are given in green. In each plot, the points that are selected for inclusion in the eVDF fit are filled with pink color. }
\label{fig_evdf_examples}
\end{figure*}
The core `c' and halo `h' electron populations are described by the sum
\begin{equation}
  f_{\rm model}\,(v_{\parallel},v_{\perp}) = f_c \,(v_{\parallel},v_{\perp}) + f_h\,(v_{\parallel},v_{\perp}),
\end{equation}
where
\begin{equation}
f_c \,(v_{\parallel},v_{\perp}) = A_{c} \;  \exp \left(-\frac{m}{2} 
    \left[] 
      \frac{(v_{\perp}-v_{\mathrm{c}\perp})^2}{T_{\mathrm{c}\perp}} +
      \frac{(v_{\parallel}-v_{\mathrm{c}\parallel})^2}{T_{\mathrm{c}\parallel}}
    \right]
  \right), 
  \end{equation}
  \begin{equation}
A_{c}  = n_{\mathrm{c}} \left(\frac{m}{2 \pi } \right)^{3/2}
  \frac{1}{T_{\mathrm{c}\perp}\sqrt{T_{\mathrm{c}\parallel}}},
\end{equation}
\begin{equation}
f_h\,(v_{\parallel},v_{\perp}) = A_{h} \;
  \left\{
    1 + \frac{m}{2\kappa-3}
    \left[
      \frac{(v_{\perp}-v_{\mathrm{h}\perp})^2}{T_{\mathrm{h}\perp}} +
      \frac{(v_{\parallel}-v_{\mathrm{h}\parallel})^2}{T_{\mathrm{h}\parallel}}
    \right]
  \right\}^{-\kappa-1},
  \end{equation}
  and
  \begin{equation}
A_{h} = n_{\mathrm{h}} 
  \left(\frac{m}{\pi (2 \kappa - 3)}
  \right)^{3/2} \frac{1}{T_{\mathrm{h}\perp}\sqrt{T_{\mathrm{h}\parallel}}}
  \frac{\Gamma(\kappa+1)}{\Gamma(\kappa-1/2)}.
\end{equation}

To reduce the number of parameters in a given fit, we apply our fits separately to
perpendicular and parallel cuts through one dimensional distribution
functions
\begin{equation}
  f_{\perp,\parallel}(v) = f_{c \perp, \parallel}(v) + f_{h \perp, \parallel}(v),
\end{equation}
where 
\begin{equation}
f_{c \perp, \parallel}(v) =  A_{c}
  \exp \left(-
      \frac{m (v-v_{\mathrm{c \perp, \parallel}})^2}{2 T_{\mathrm{c \perp, \parallel}}}
  \right),
\end{equation}
\begin{equation}
f_{h \perp, \parallel}(v) =  A_{h} \;  \left[
    1 + \frac{m}{2\kappa-3}
      \frac{(v-v_{h \perp, \parallel})^2}{T_{h \perp, \parallel}}
  \right]^{-\kappa-1},
\end{equation}
where the parallel and perpendicular subscripts have been eliminated.
We use a common form for both parallel and perpendicular cuts:
\begin{equation}
  f  =  A_0 \exp\left( - \frac{(v - A_1)^2}{A_2} \right) 
  + A_3 \left\{ 1 + \left[ \frac{(v - A_4)^2}{A_5} \right]^{-A_6} \right\} 
\end{equation}
In both the perpendicular and parallel directions, there are seven fit parameters. Starting each fit with a reasonable initial guess  improves the speed and stability of the algorithm. For the perpendicular case, the fit is initialized using the TNR measurements of the core and halo density and temperature. The starting guess for the $\kappa$ parameter is based on solar wind speed ($\kappa = 6$ if $V_{sw} < 500$\,km/s, $\kappa = 5$ if $500< V_{sw} < 650$\,km/s), and $\kappa = 4$ if $V_{sw} > 650$\,km/s) from previous experience.
The drift velocity parameters $A_1$ and $A_4$ are fixed at zero (consistent with the gyrotropic assumption), reducing the number of fit parameters to 5.

The perpendicular fit determines perpendicular temperature (from $A_2$ and $A_5$), the $\kappa$ value from $A_6$, and the overall phase space amplitude of the core and halo populations $A_0$ and $A_3$. 
The parallel fit is performed after the perpendicular fit, keeping $A_0$, $A_3$, and $A_6$ constant. The parallel fit determines the core and halo drift speed $A_1$ and $A_4$, and the parallel temperatures (from $A_2$ and $A_5$).
Using the measured amplitudes $A_0$ and $A_3$ and the measured temperature and $\kappa$, the core and halo densities can be determined.

Figure \ref{fig_evdf_examples} shows two typical eVDFs in the solar wind at 1\,au measured by the Wind/3DP electrostratic analyzers EESA-L and EESA-H. The left panels (a) and (c) show an eVDF in the slow solar wind (at 1995-06-19/00:06:38), and the right panels (b) and (d) show an eVDF in the fast solar wind (at 1995-06-19/23:13:59). The top panels (a) and (b) show cuts through the eVDF in one of the two directions perpendicular to the local magnetic field {\bf B}: the diamonds are data points from EESA-L and the asterisks from EESA-H. The dotted lines represent the one-count level for EESA-L and EESA-H.  The blue dashed line in Figs. \ref{fig_evdf_examples}a and \ref{fig_evdf_examples}b represents the sum of Maxwellian and Kappa distributions calculated using the QTN fit parameters (indicated in blue).  The red line in Fig.~\ref{fig_evdf_examples}a and \ref{fig_evdf_examples}b represents the fit to the measured perpendicular eVDF cut; the resulting fit parameters are indicated in red.  The bottom panels (c) and (d) show cuts through the eVDF in the direction parallel to {\bf B}. The perpendicular fit is reported in red, and the perpendicular fit parameters are used to initialize the parallel eVDF fit.

As shown in Figs. \ref{fig_evdf_examples}c and \ref{fig_evdf_examples}d, the points that are selected for inclusion in the eVDF parallel fit are filled with pink color. Points that are determined by the algorithm to represent the strahl and points close to the one-count level are not included in the eVDF fit. The result of the parallel fit is indicated in green (green curves and green parameters). The solar wind speed and s/c potential retained by the algorithm are given on the top of each panels, as well as a quality measure of the nonlinear least-square fit (see section 3.7 below) and its $\chi^2$.

\subsection{Extracting the electron Strahl}\label{extr_strahl}

\begin{figure*}[h]
 \centering
 \includegraphics[width=0.8\hsize]{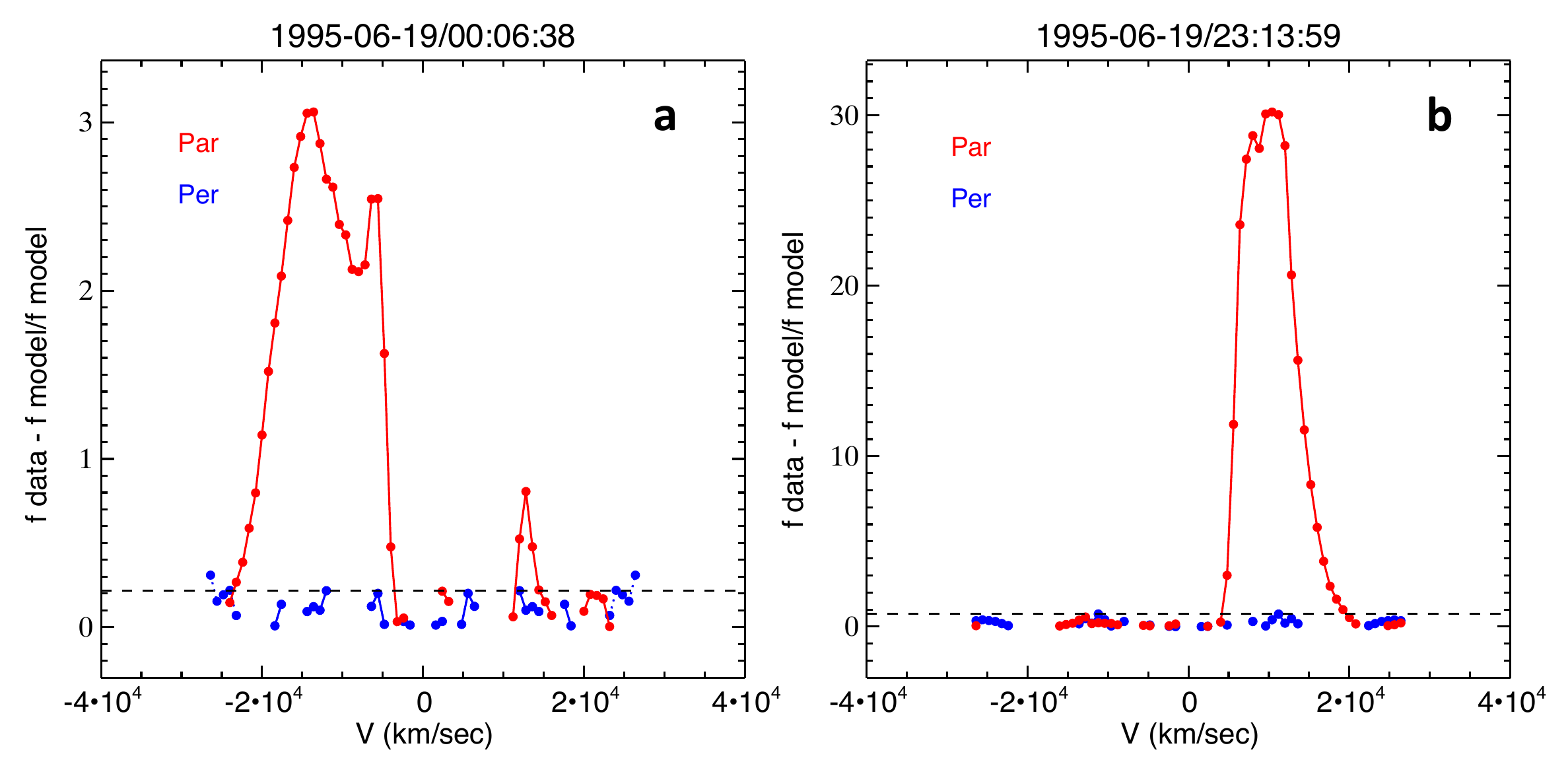}
\caption{\small This figure illustrates the algorithm used to extract the strahl from the eVDFs of Fig. \ref{fig_evdf_examples}; the left panel (a) shows the slow wind eVDF and the right panel (b) shows the fast wind eVDF.  Cuts through the $\Delta$ distribution (see text) in the parallel and perpendicular direction are indicated in red and blue respectively. The obvious peak in the parallel cut in red shows the range and structure of the strahl.  In the fast wind, the strahl is much more prominent than in the slow wind. However, the strahl is wider in the slow wind than in the fast wind. The dashed horizontal line represents the threshold for the automated extraction of the strahl.}
\label{fig_delta}
\end{figure*}
After the final perpendicular and parallel eVDF fit described in Fig. \ref{fig_evdf_examples}, $f_{\rm{model}}(v_{\perp},v_{\parallel})$ is the 2D distribution function constructed from the perpendicular and parallel fit parameters, not including strahl, i.e., a {\it core-halo} model.

\begin{figure*}
 \centering
 \includegraphics[width=0.8\hsize]{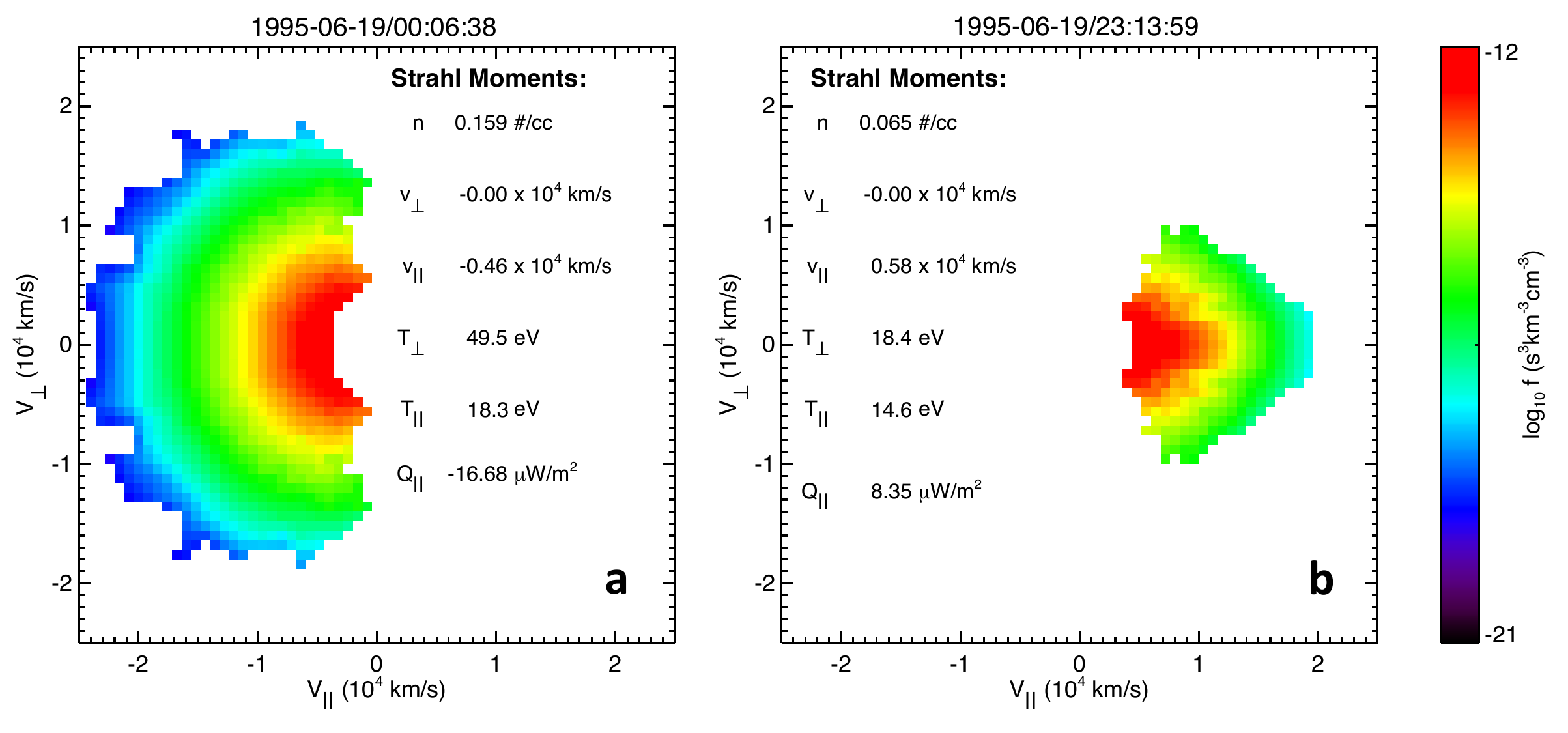}
\caption{\small The 2D distribution of the strahl after extraction from the total observed eVDF. The left panel (a) and the right panel (b) are the same slow and fast solar wind examples as in Figs. \ref{fig_evdf_examples} and \ref{fig_delta}. The calculated strahl moments for each example are listed.}
\label{fig_strahl2D}
\end{figure*}
The combined EESA-L and EESA-H distributions encompass over 10 orders of magnitude in phase space density. Therefore, detection of a feature such as the strahl, in comparison to a difference between the measured distribution $f_{\rm{data}}$ and the model $f_{\rm{model}}$ due to ``noise'' requires use of a normalized distribution.
We define the $\Delta$ distribution as the normalized difference between $f_{\rm{data}}$ and  $f_{\rm{model}}$:
\begin{equation}
\Delta (v_{\perp},v_{\parallel}) = \frac{f_{\rm{data}}(v_{\perp},v_{\parallel}) - f_{\rm{model}}(v_{\perp},v_{\parallel})}{f_{\rm{model}}(v_{\perp},v_{\parallel})}.
\end{equation}

From  $\Delta (v_{\perp},v_{\parallel})$, we extract perpendicular and parallel cuts. Values of the perpendicular cut, $\Delta (v_{\perp},v_{\parallel}=0)$, are indicated in blue in Fig \ref{fig_delta} and are typically small (less than 1) because the strahl is not present in the perpendicular cut.  In contrast, values of the parallel cut, $\Delta (v_{\perp} =0,v_{\parallel})$, indicated in red in Fig. \ref{fig_delta}, can reach high values, revealing clearly the strahl and its shape.
The strahl structures shown in Figure \ref{fig_delta} are typical of those observed by Wind in the unperturbed solar wind: the strahl is uni-directional with a peak density (relative to the core/halo density) at a parallel speed between $1\times 10^4$ and 2$\times10^4$\,km/s, and is limited in velocity/energy with a cutoff at usually less than $4\times 10^4$\,km/s.

The dashed horizontal line in Fig.~\ref{fig_delta} represents a threshold for the automated extraction of the strahl. For each eVDF, this threshold is determined using the perpendicular cut as a control (assuming that the perpendicular cut contains no strahl): values of the $\Delta$-distribution in the parallel direction which have values significantly greater than typical values in the perpendicular cut are counted as strahl.
The obvious peak in the parallel cut in red shows the range and structure of the strahl.  In the fast wind, the strahl is much more prominent than in the slow wind. Indeed, $\Delta \sim 3$ for the slow wind example (left panel a) and $\Delta \sim 30$ for the fast wind example (right panel b), i.e. 10 times higher.  However, the strahl is wider in the slow wind than in the fast wind.

To extract the 2D strahl distribution from the measured $f_{\rm{data}}$, we select all data points characterized by $\Delta > \delta_0$, where $\delta_0$ is a set threshold, and define the strahl distribution as the difference between the measured eVDF and the {\it core-halo} model:
\begin{equation} 
f_s (v_{\perp},v_{\parallel}) = f_{\rm{data}}(v_{\perp},v_{\parallel}) - f_{\rm{model}}(v_{\perp},v_{\parallel}) \;\;\;\;\;\;\; \rm{for} \;\;\; \Delta > \delta_0
\end{equation}

Once the 2D strahl distribution is extracted, we characterize it by integrating strahl moments, namely density, bulk speed (in the solar wind frame), intrinsic parallel and perpendicular temperatures, as well as heat flux.  Fig. \ref{fig_strahl2D} shows the 2D strahl distribution for the above examples of slow and fast wind (left -a- and right -b- panels respectively). The strahl moments are also given.

In these representative examples, the strahl is  not only wider in energy range but also broader in pitch angle in the slow wind compared to the fast wind.  The temperature anisotropy of the strahl, determined from the intrinsic parallel and perpendicular temperatures (i.e., calculated in its own frame) yield a good a characterization of its angular width. In the example of Fig. \ref{fig_strahl2D}, the temperature anisotropy is 1.26 for the fast wind (right panel) and 2.70 for the slow wind (left panel), confirming that the strahl is broader in the slow wind compared to the fast wind.
 
 \subsection{Total electron moments and heat flux}

The eVDF fit process described above yields independent parameters of the core, halo and strahl populations for each measured, processed and corrected distribution function. In addition to the core, halo and strahl thus determined, we also integrate each full and calibrated eVDF up to the energies of the super halo to determine the total electron density $n_e$, parallel and perpendicular temperatures $T_{e\parallel}$ and $T_{e\perp}$, as well as the electron heat flux ${\bf Q}_e$.  By construction, this total heat flux is simply the parallel heat flux ${\bf Q}_e = Q_{e\parallel} \hat{b}$, since the eVDF is symmetric in the perpendicular direction within the gyrotropy assumption.    
During this process, we verify that $n_e$ thus determined is  consistent with $n_c + n_h + n_s$ as determined above. 

\subsection{Quality assessment of eVDF fits}

The quality of each eVDF fit is evaluated once the fit procedure is complete. The quality factor is a numerical measure with values between 0 (worst) and 10 (best), based on the convergence of the fit algorithm, the reported $\chi^2$ value of the parallel and perpendicular fits, the agreement between EESA-L and EESA-H over their common energy range, sufficient density of angular bins in the vicinity of the strahl, and independently measured quality of the QTN measurements used to initialize the fit. Times when the EESA detectors are not in a suitable mode for analysis are also eliminated (for example, during certain intervals the integration time for EESA-H is set too low to accumulate sufficient statistical counts in the strahl energy range). The remainder of this paper is based on 280\,000 full eVDF fits which were determined to be sufficiently high quality.

\section{Results}

The  application of the algorithm layed out in Section~\ref{evdf_processing} to our 280\,000 eVDFs creates a large dataset of plasma parameters for the total electron species and the core, halo and strahl populations. In this section, we present a first statistical analysis of these parameters.  The underlying statistics are also summarized in the Appendix.

\begin{figure}[ht]
\centering
\includegraphics[width=0.8\hsize]{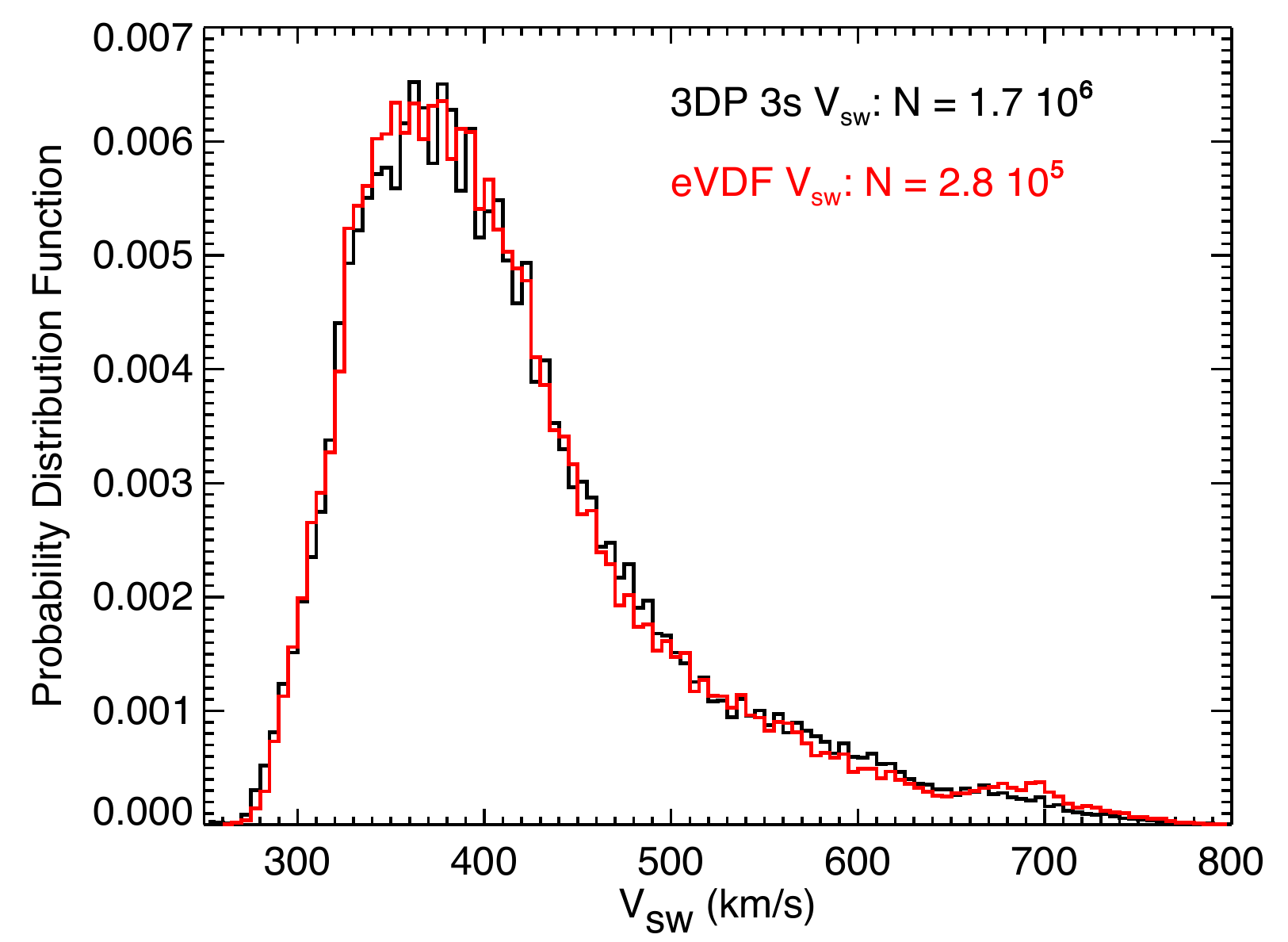}
\caption{Probability distributions of the solar wind speed $V_{sw}$. The black histogram represents the distribution based on our eVDF analysis, and the red histogram represents the distribution based on 3DP onboard proton moments, normalized to the maximum of the eVDF historgram. }
\label{histo_sample}
\end{figure}

Figure~\ref{histo_sample} shows the probability distribution of solar wind bulk speeds in our full dataset. The black histogram shows the distribution of $V_{sw}$ resulting from our eVDF fits. The binning in solar wind speed uses a binsize of 5\,km/s between 250 and 800\,km/s. 
The red histogram shows the distribution of the same parameter, $V_{sw}$, sampled from the original 3-second proton 3DP data (onboard moments) from the same 4-year time period, normalized to the maximum of the eVDF histogram. The proton histogram is based on 1.7 millions data points. The comparison between the electron histogram and the proton histogram shows that our eVDF processing does not introduce any bias in our sampled distribution of solar-wind speeds compared to the 3DP proton data.

For the statistical analysis of solar wind electron data at 1\,au, we now summarize our data analysis in column-normalized 2D histograms of electron parameters from the integration of the final combined (EESA-L and EESA-H) eVDFs, corrected for spacecraft-potential effects in Figures~\ref{fig_totalparam1} through~\ref{fig_chs_ratios}. We choose the solar wind speed as a reliable statistical ordering parameter for the histograms in Figures~\ref{fig_totalparam1} through~\ref{fig_tratios}. We apply the same bin widths to the following histograms as in Figure~ \ref{histo_sample}, i.e., a binsize of 5\,km/s. The color bar represents the logarithm of the number of counts per bin, normalized to the maximum number of counts per bin in each column. The black circles mark the mean in each column, and the black vertical lines represent the standard deviation based on a Gaussian fit to the normalized values in each column (see Appendix).

\subsection{Statistics of the total electron distribution}

\begin{figure*}[ht]
\centering
\includegraphics[width=0.9\hsize]{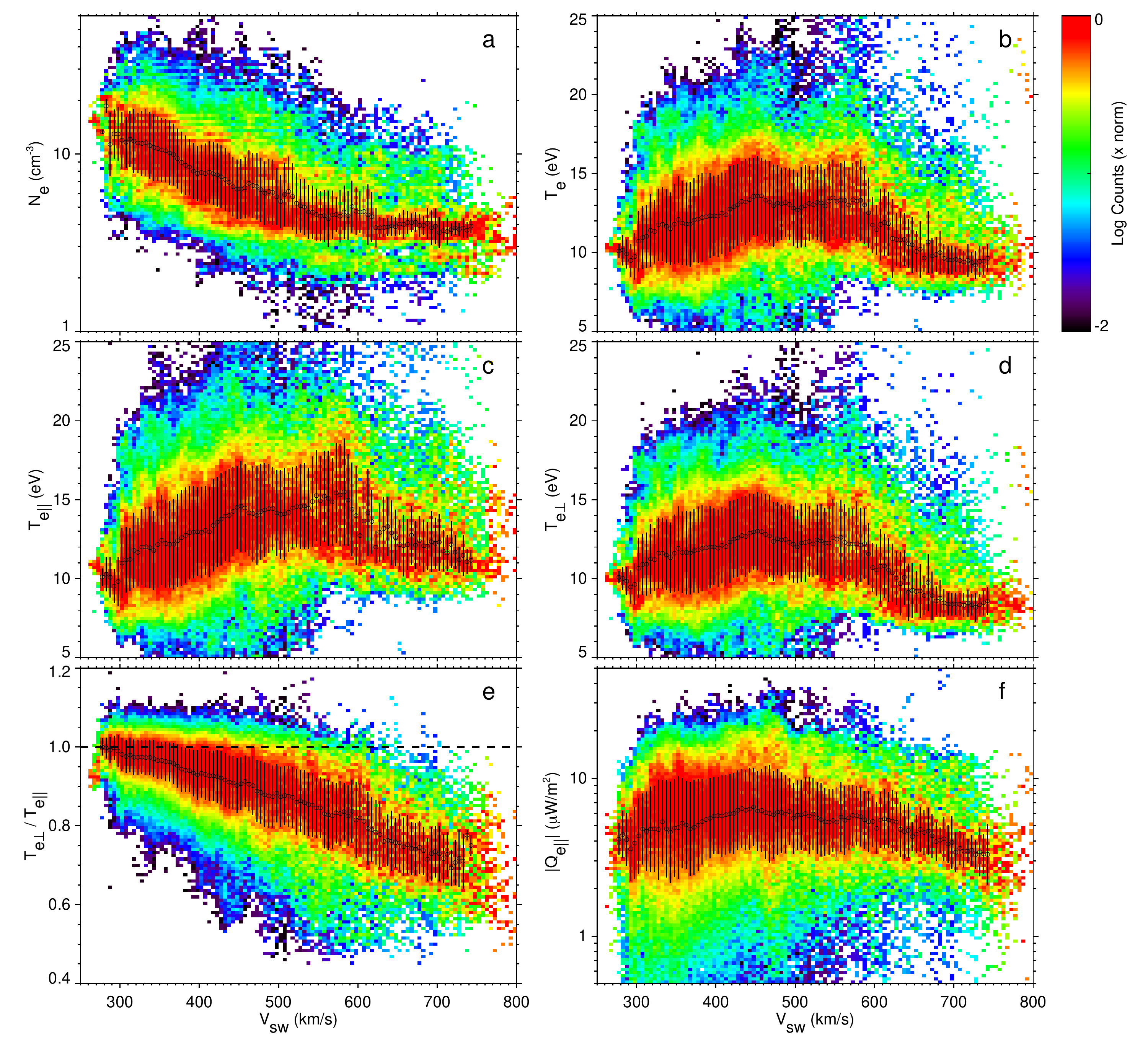}
\caption{Column-normalized 2D histrograms of parameters of the total electron species: (a) electron density $N_e$, (b) total electron temperature $T_e$, (c) temperature $T_{e\parallel}$ of the total electron distribution in the direction parallel to the mean magnetic field, (d) temperature $T_{e\perp}$ of the total electron distribution in the direction perpendicular to the mean magnetic field, (e) temperature ratio $T_{e\perp}/T_{e\parallel}$ of the total electron distribution, and (f) magnitude of the parallel heat flux $|Q_{e\parallel}|$ of the total electron distribution.}
\label{fig_totalparam1}
\end{figure*}

Figure~\ref{fig_totalparam1} displays histograms of the moments of the total electron distribution function, not separated by core, halo, and strahl.  Panel (a) shows the distribution of the total electron density, $N_e$. The average $N_e$ drops from $\sim 13\,\mathrm{cm}^{-3}$ in slow wind to $\sim 4\,\mathrm{cm}^{-3}$ in fast wind. Also the variability in $N_{\mathrm e}$ decreases with increasing $V_{sw}$. Panel (b) shows the total electron temperature $T_e$. Although the analysis suggests some variation with wind speed, the statistical trend is less pronounced than the trend in $N_{e}$. For most cases, the total electron temperature varies between $\sim 8\,\mathrm{eV}$ and $\sim 16\,\mathrm{eV}$. However, we also see a larger variation in $T_{e}$ at smaller $V_{sw}$ than at large $V_{sw}$. Panels (c) and (d) show the temperatures $T_{e \perp }$ and $T_{e \parallel}$ of the total electron population perpendicular and parallel with respect to the background magnetic field, respectively. Like in the case of the total $T_{e}$, $T_{e\parallel}$ and $T_{e\perp}$ show more variation in fast wind than in slow wind. We especially see that $T_{e\perp}$ assumes smaller average values in the fast wind ($\sim 8\,\mathrm{eV}$) compared to the slow wind ($\sim 12\,\mathrm{eV}$). Panel (e) confirms this statistical trend by showing the histogram for the ratio $T_{e\perp}/T_{e\parallel}$. The total eVDF is less anisotropic in slow wind and exhibits an average anisotropy of $T_{e\perp}/T_{e\parallel}\sim 0.7$ in fast wind. Panel (f) shows the magnitude of the parallel heat flux $|Q_{e\parallel}|$ of the total electron heat flux.  $|Q_{e\parallel}|$ exhibits more variation and a slightly higher value in slow solar wind ($\sim 4\,\mu\mathrm W/\mathrm m^2$) than in fast solar wind ($\sim 3\,\mu\mathrm W/\mathrm m^2$).

\begin{figure*}[ht]
\centering
\includegraphics[width=0.9\hsize]{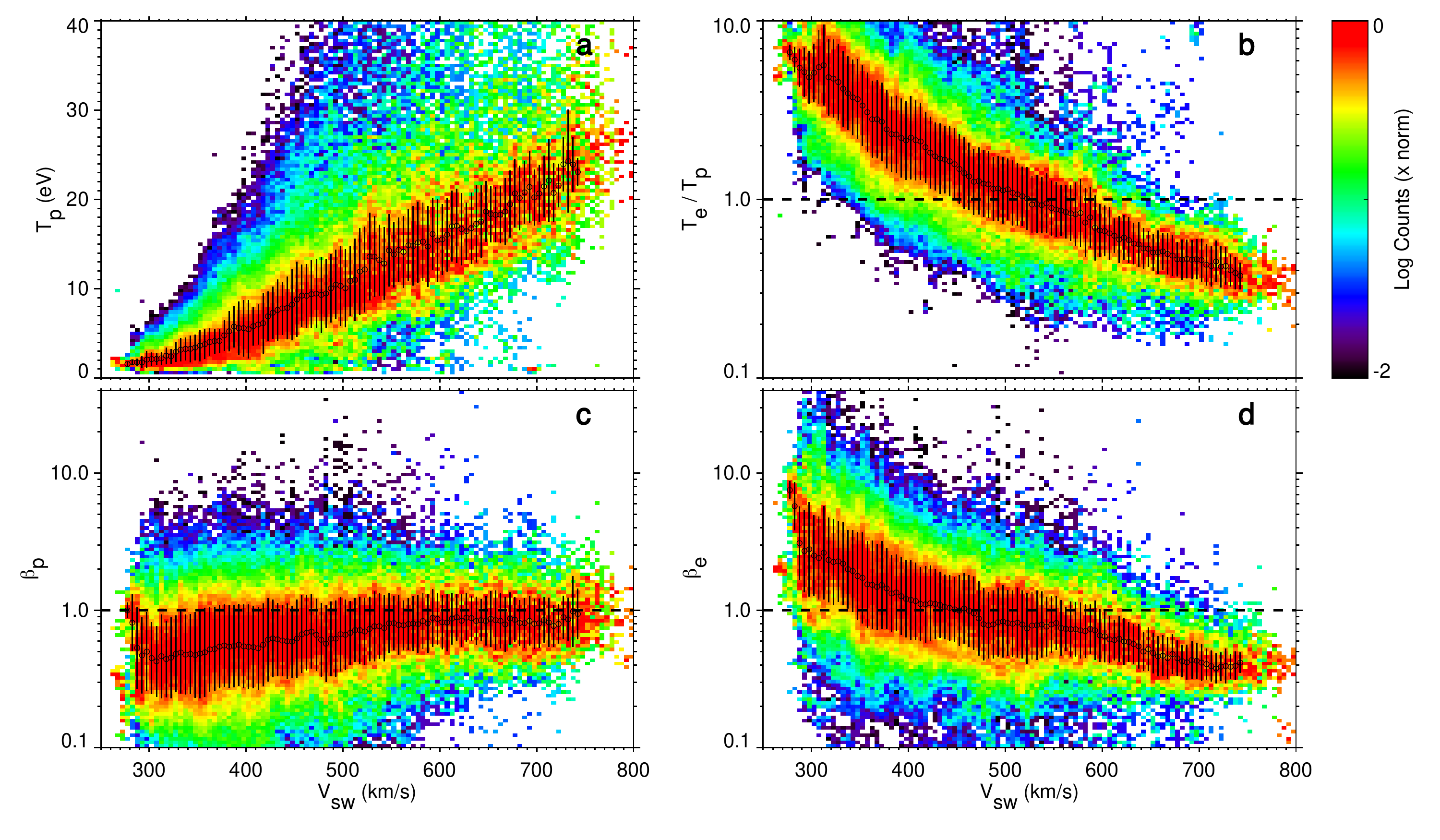}
\caption{Column-normalized 2D histrograms of proton parameters and derived total electron parameters: (a) total proton temperature $T_p$, (b) temperature ratio $T_{e}/T_p$ of the total proton and electron distributions, (c) $\beta_p$ of the total proton distribution, and (d) $\beta_e$ of the total electron distribution.}
\label{fig_totalparam2}
\end{figure*}

Figure~\ref{fig_totalparam2} shows the statistical results for the proton data and derived parameters for the total eVDFs in our dataset. Panel (a) shows that the proton temperature follows a clear positive trend with $V_{sw}$, which is a well-known property of the solar wind \citep{Burlaga:1970, Burlaga:1973, Lopez:1986}. The temperature increases from $\sim 2\,\mathrm{eV}$ in slow wind to $\sim 24\,\mathrm{eV}$ in fast wind. This increase in $T_{p}$, combined with the $V_{sw}$-dependence of $T_e$ shown in Figure~\ref{fig_totalparam1} (b), explains the strong anti-correlation between $T_e/T_p$ and $V_{sw}$ seen in panel (b). The total proton and electron temperatures are on average approximately equal for $V_{sw}\sim 530\,\mathrm{km/s}$. Panels (c) and (d) show the ratios between the total proton and electron thermal energy densities and the magnetic-field energy density, respectively, which we calculate as $\beta_p=8\pi n_pk_BT_p/B^2$ and $\beta_e=8\pi n_ek_BT_e/B^2$, where $k_B$ is the Boltzmann constant. We see that $\beta_p\lesssim 1$ on average with only a slight dependency on $V_{sw}$, while $\beta_e$ shows, on average, a clear anti-correlation with $V_{sw}$. We find that $\beta_e\sim 1$ for $V_{sw}\sim 450\,\mathrm{km/s}$. We note that fast wind exhibits less variability in both $\beta_p$ and $\beta_e$ compared with slow wind.

\subsection{Statistics of core, halo and strahl components}

\begin{figure*}[ht]
\centering
\includegraphics[width=0.9\hsize]{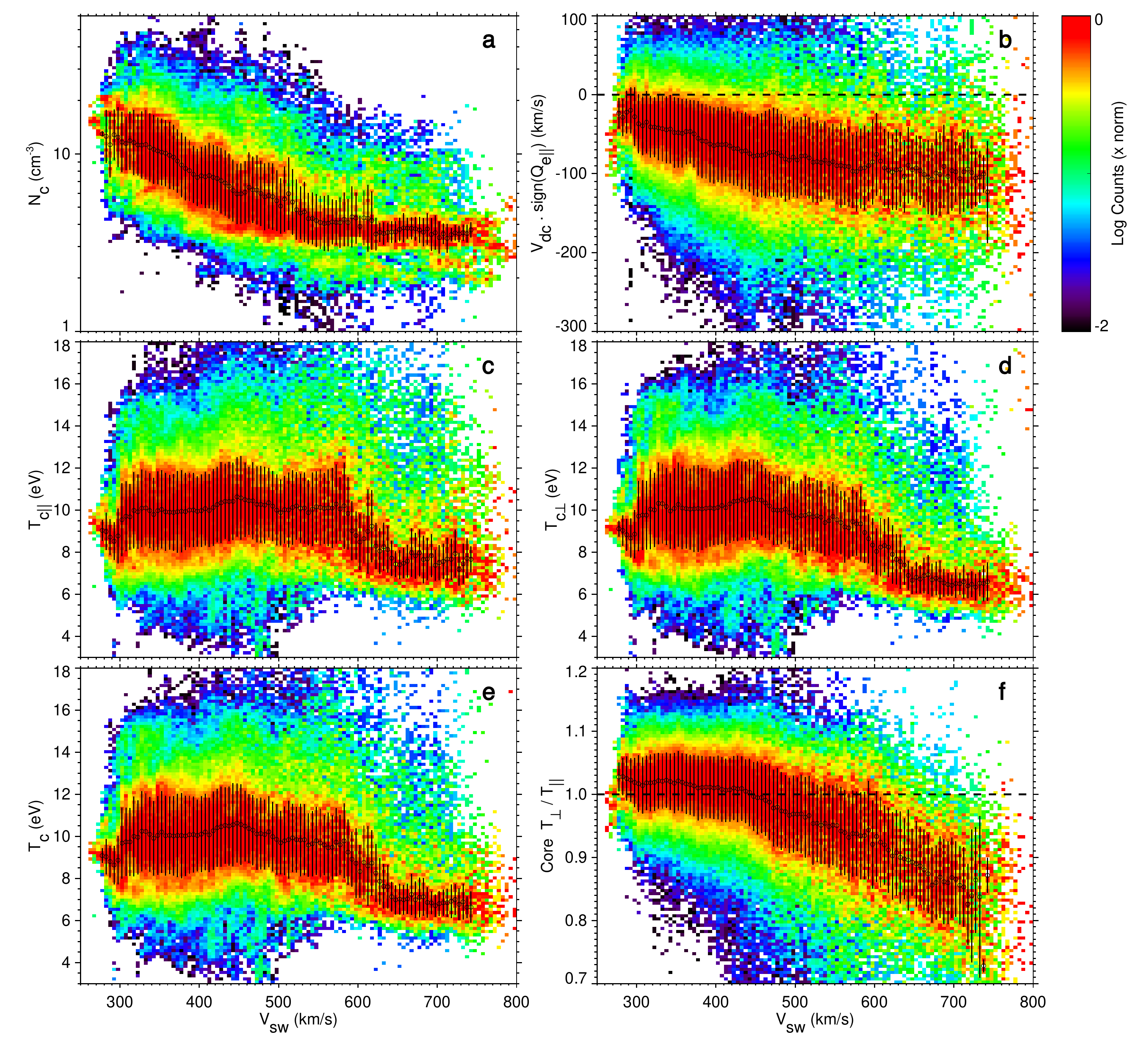}
\caption{Column-normalized 2D histrograms of electron core parameters: (a) core density $N_c$, (b) sign-corrected drift speed $V_{dc}\,\mathrm{sign}(Q_{e\parallel})$ between core and proton bulk velocity, (c) temperature $T_{c\parallel}$  of the core in the direction parallel to the mean magnetic field, (d) temperature $T_{c\perp}$  of the core in the direction perpendicular to the mean magnetic field, (e) total temperature $T_c$ of the core, and (f) temperature ratio $T_{\perp}/T_{\parallel}$ of the core.}
\label{fig_corestats}
\end{figure*}

Figure~\ref{fig_corestats} shows column-normalized histograms for the electron core parameters as functions of $V_{sw}$. Panel (a) shows the core number density $N_c$. On average, $N_c$ varies from $\sim 12\,\mathrm{cm}^{-3}$ in slow solar wind to $\sim 4\,\mathrm{cm}^{-3}$ in fast solar wind. Likewise, the variability in $N_c$ decreases with increasing $V_{sw}$. Panel (b) shows the drift speed $V_{dc}$ between the electron core and the protons, multiplied with the sign of $Q_{e\parallel}$. By multiplying $V_{dc}$ with $\mathrm{sign(Q_{e\parallel})}$, we correct for the sign ambiguity due to the choice of our coordinate system. If $V_{dc}\,\mathrm{sign(Q_{e\parallel})}<0$, the core bulk velocity is directed towards the Sun in the coordinate system centered on the proton bulk velocity, and vice versa. We find indeed that  $V_{dc}\,\mathrm{sign(Q_{e\parallel})}<0$. In the slow wind, $|V_{dc}|$ is smaller ($\sim 20\,\mathrm{km/s}$) than in the fast wind ($\sim 100\,\mathrm{km/s}$). Panels (c) and (d) show the core temperatures in the directions parallel ($T_{c\parallel}$) and perpendicular ($T_{c\perp}$) with respect to the magnetic field. Both temperatures are greater in the slow solar wind ($T_{c\parallel}\sim T_{c\perp}\sim 10\,\mathrm{eV}$) than in the fast solar wind ($T_{c\parallel}\sim 8\,\mathrm{eV}$, $T_{c\perp}\sim 7\,\mathrm{eV}$). The scalar core temperature $T_c$ reflects the same trend, as shown in panel (e). Consistent with the behavior of $T_{c\parallel}$ and $T_{c\perp}$,  the temperature ratio $T_{\perp}/T_{\parallel}$ of the core decreases from isotropy in slow wind to a value of $\sim 0.83$ in fast wind.

\begin{figure*}[ht]
\centering
\includegraphics[width=0.9\hsize]{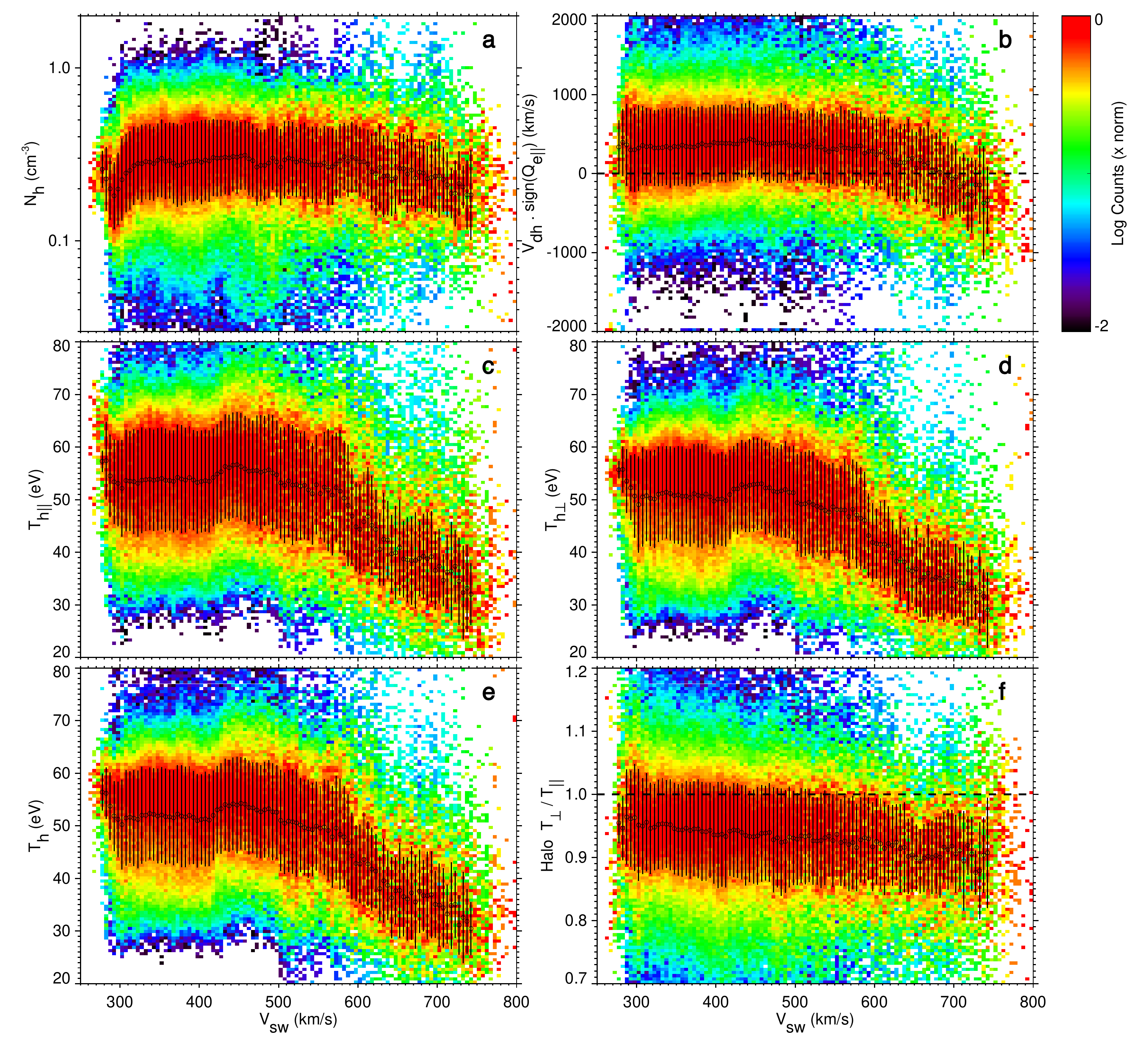}
\caption{The same as Figure~\ref{fig_corestats}, but for the halo instead of the core.}
\label{fig_halostats}
\end{figure*}

Figure~\ref{fig_halostats} shows the same histograms as Figure~\ref{fig_corestats}, but for the halo instead of the core. Panel (a) shows the halo density $N_h$, which is mostly independent of $V_{sw}$ with a value of $\sim 0.3\,\mathrm{cm}^{-3}$. The halo drift $V_{dh}$ is directed away from the Sun and assumes values of $\sim 400\,\mathrm{km/s}$ in slow solar wind and almost vanishes in fast solar wind. There is evidence for a Sunward halo drift at $V_{sw}\gtrsim 700\,\mathrm{km/s}$; however, the statistics for $V_{dc}$ in this regime is less reliable than at smaller $V_{sw}$. A visual inspection of the individual fits (like those shown in \ref{fig_evdf_examples}) confirms the direction of the drifts -- sunward core and halo drifts -- and the goodness of the fit overall. Panels (c) and (d) show the halo temperatures in the directions parallel ($T_{h\parallel}$) and perpendicular ($T_{h\perp}$) with respect to the magnetic field. Both temperatures are approximately constant ($T_{h\parallel}\sim 53\,\mathrm{eV}$ and $T_{h\perp}\sim 50\,\mathrm{eV}$) in slow solar wind and drop to $T_{h\parallel}\sim 32\,\mathrm{eV}$ and $T_{h\perp}\sim 30\,\mathrm{eV}$ in fast wind. The constant-temperature regime occurs at $V_{sw}\lesssim 600\,\mathrm{km/s}$. The scalar halo temperature also follows this trend from $\sim50\,\mathrm{eV}$ in slow wind to $\sim 32\,\mathrm{eV}$ in fast wind. The ratio $T_{\perp}/T_{\parallel}$ of the halo is on average less than one, with a slight inverse trend with $V_{sw}$ on average. We note, however, that there is a large variation in $T_{\perp}/T_{\parallel}$ of the halo, so that there are significant times with $T_{\perp}/T_{\parallel}\gtrsim 1$, especially during slow wind times.

\begin{figure*}[ht]
\centering
\includegraphics[width=0.9\hsize]{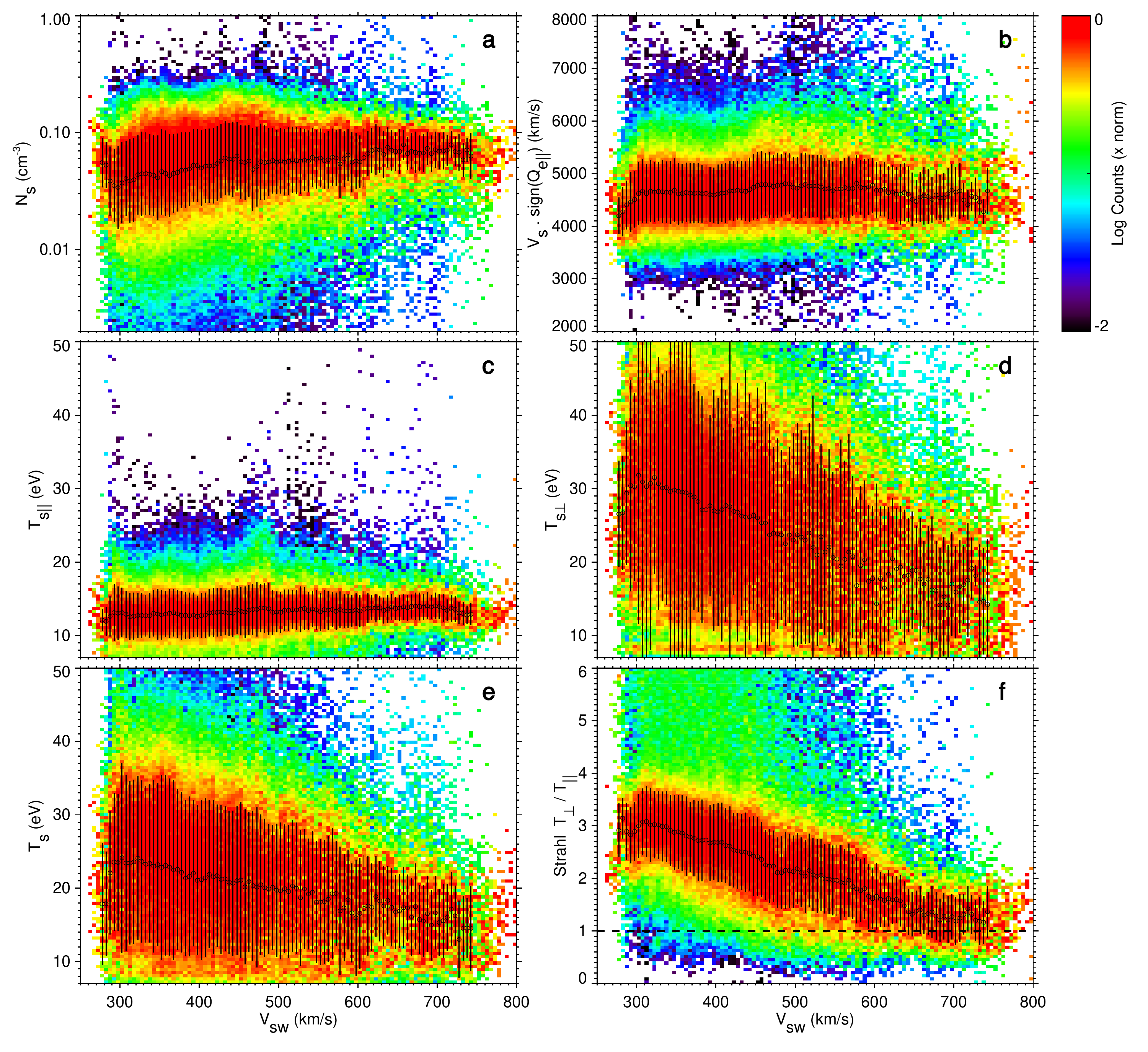}
\caption{The same as Figure~\ref{fig_corestats}, but for the strahl instead of the core.}
\label{fig_strahlstats}
\end{figure*}

Figure~\ref{fig_strahlstats} shows the same histograms as Figure~\ref{fig_corestats}, but for the strahl instead of the core. Panel (a) shows the strahl density $N_s$, which varies from $\sim 0.04\,\mathrm{cm}^{-3}$ in slow wind to $\sim 0.08\,\mathrm{cm}^{-3}$ in fast wind. The slow wind shows more variability in $N_s$ than the fast wind. Panel (b) shows the drift speed $V_{s}$ of the strahl multiplied by $\mathrm{sign}(Q_{e\parallel})$. The strahl is directed away from the Sun and exhibits an approximately constant average drift speed of $\sim 4700\,\mathrm{km/s}$ in the proton frame. Also the width of the  column-normalized data distribution is mostly independent of $V_{sw}$. Panel (c) shows the strahl temperature $T_{s\parallel}$ in the direction parallel to the mean magnetic field. $T_{s\parallel}\sim 14\,\mathrm{eV}$ on average and is mostly constant with $V_{sw}$. On the contrary, the strahl temperature $T_{s\perp}$ in the direction perpendicular to the mean magnetic field shows a large variation and a strong anti-correlation with $V_{sw}$ on average (see panel (d)). While the average $T_{s\perp}\sim 30\,\mathrm{eV}$ in slow wind, it is $\sim 15\,\mathrm{eV}$ in fast wind. Consequently, the scalar strahl temperature $T_s$, shown in panel (e), also decreases with increasing $V_{sw}$ from $\sim 24\,\mathrm{eV}$ in slow wind to $\sim 15\,\mathrm{eV}$ in fast wind. Panel (f) illustrates that the temperature ratio $T_{\perp}/T_{\parallel}$ of the strahl varies from $\sim 3$ in the slow solar wind to $\sim 1.4$ in fast wind. As mentioned in Section~\ref{extr_strahl}, $T_{s\perp}/T_{s\parallel}$ is a good parameter to characterize the angular width of the strahl, and this angular width of the strahl displays a linear anti-correlation with solar wind speed at 1\,au.  

\subsection{Current balance}

\begin{figure}[ht]
\centering
\includegraphics[width=0.9\hsize]{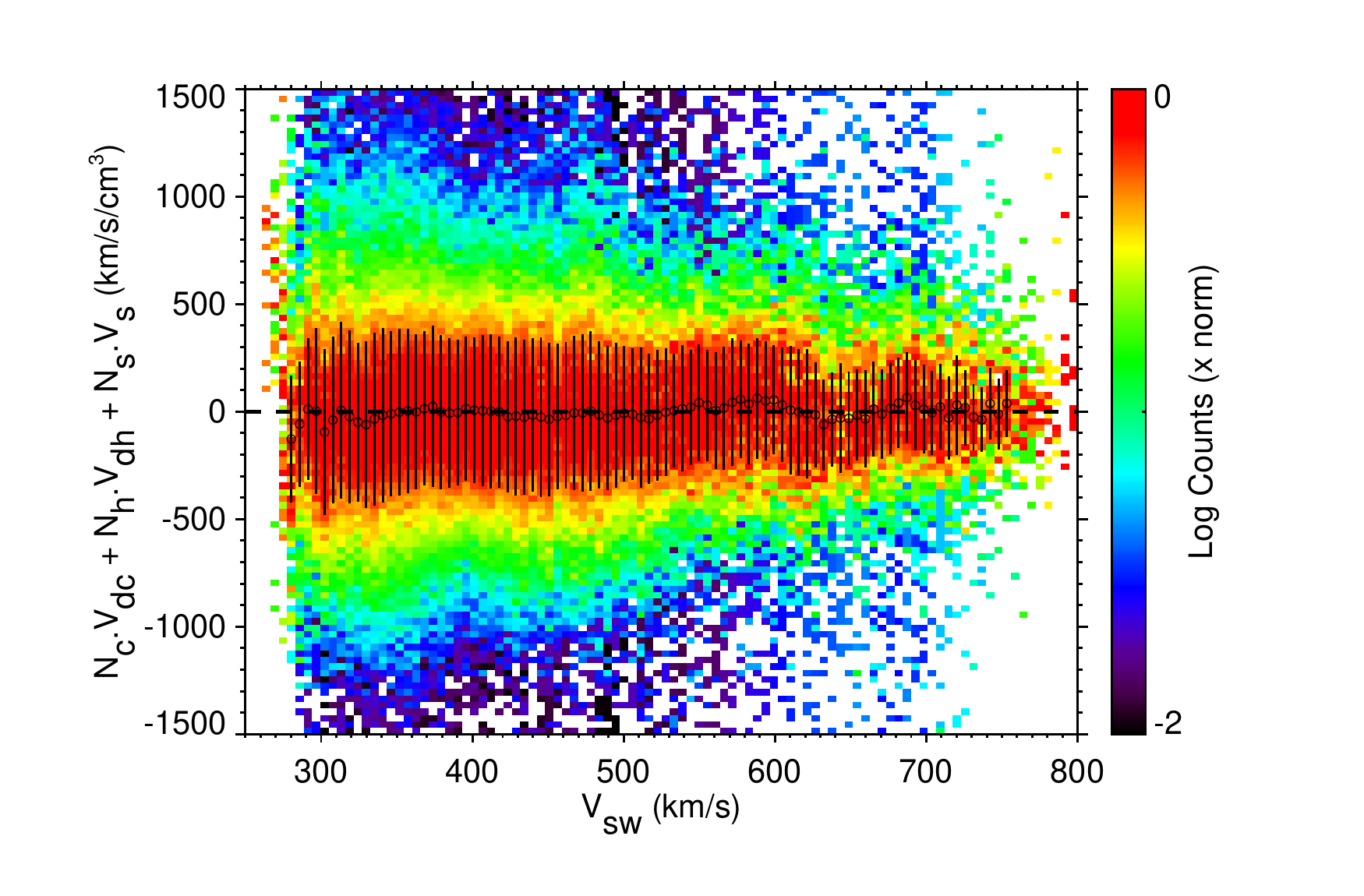}
\caption{Current balance of the total electron distribution function.}
\label{fig_currentbalance}
\end{figure}

As an independent check of the density and drift-speed measurements, we analyze the current-balance condition of our fit results for the eVDF in Figure~\ref{fig_currentbalance}. We show a column-normalized histogram of the quantity $N_c\,V_{dc}+N_h\,V_{dh}+N_s\,V_s$ as a function of $V_{sw}$. In the proton rest frame, current balance is fulfilled when this quantity is zero. As expected, the results in Figure~\ref{fig_currentbalance} are consistent with current balance.

\subsection{Intercorrelations between electron parameters}

Since most of the parameters shown thus far exhibit large variability in our histograms, it is not directly possible to deduce ratios between these parameters from the histograms. Instead, the ratios between electron parameters can follow largely independent trends with $V_{sw}$. In this section, we analyze some of these ratios to study intercorrelations between a selection of electron parameters.

\begin{figure}[ht!]
\centering
\includegraphics[width=0.9\hsize]{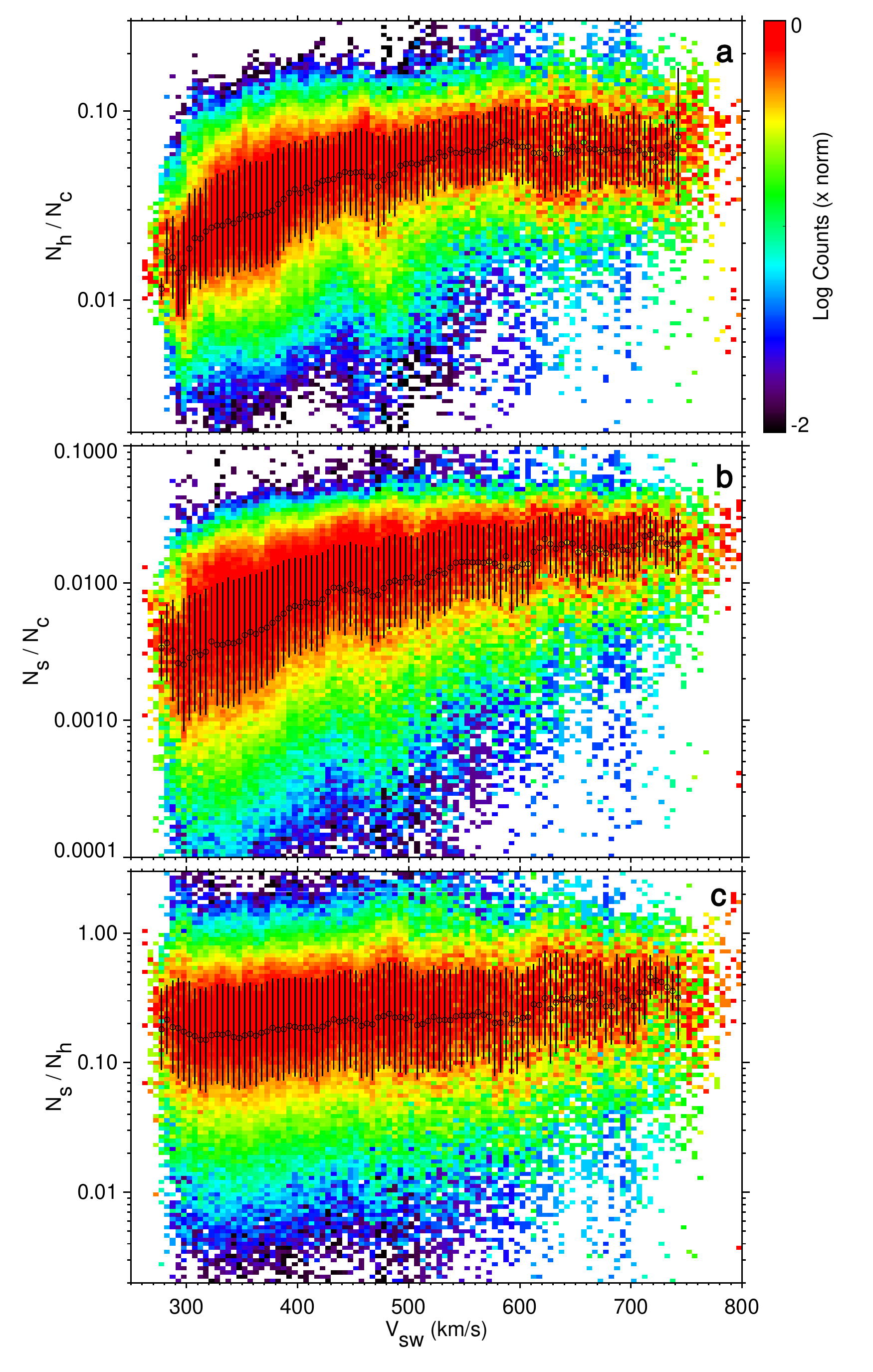}
\caption{Density ratios of the electron components as functions of $V_{sw}$. (a) $N_h/N_c$, (b) $N_s/N_c$, and (c) $N_s/N_h$.}
\label{fig_nratios}
\end{figure}

Figure~\ref{fig_nratios} shows the ratios between the densities of the three electron components from our fit results as functions of $V_{sw}$. Panel (a) shows the ratio between the halo density $N_h$ and the core density $N_c$. We find an increase in $N_h/N_c$ with $V_{sw}$ from $\sim 0.02$ in slow wind to $\sim 0.08$ in fast wind. Panel (b) shows the ratio between the strahl density $N_s$ and the core density $N_c$. It increases with $V_{sw}$ from $\sim0.003$ in slow wind to $\sim 0.02$ in fast wind. Panel (c) displays the ratio between the strahl density $N_s$ and the halo density $N_h$. Similar to the ratios shown in panels (a) and (b), this ratio increases with $V_{sw}$ from $\sim 0.2$ in slow wind to $\sim 0.4$ in fast wind.

\begin{figure}[ht!]
\centering
\includegraphics[width=0.9\hsize]{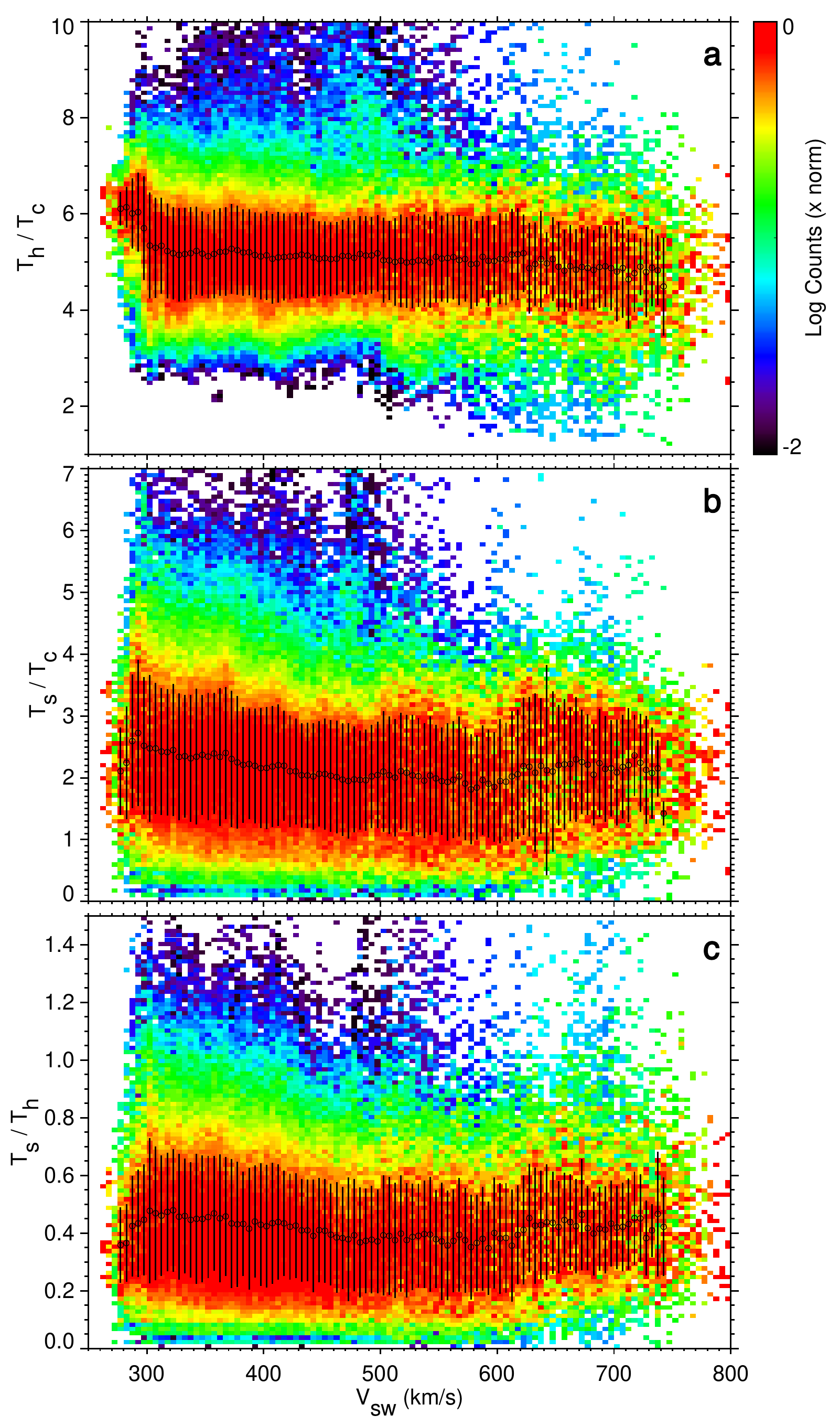}
\caption{Ratios of the scalar temperatures of the electron components as functions of $V_{sw}$. (a) $T_h/T_c$, (b) $T_s/T_c$, and (c) $T_s/T_h$.}
\label{fig_tratios}
\end{figure}

Figure~\ref{fig_tratios} shows ratios between the scalar temperatures of the electron core, halo, and strahl populations as functions of $V_{sw}$. Panel (a) shows the ratio between the scalar halo temperature $T_h$ and the scalar core temperature $T_c$. This ratio is largely independent of $V_{sw}$ and assumes a value of $\sim 5$ on average. We note that there is a slight increase in $T_h/T_c$ to $\sim 6$ in the narrow regime with $V_{sw}\lesssim 300\,\mathrm{km/s}$. Outside of this regime, the variability of $T_h/T_c$ is mostly independent of $V_{sw}$. Panel (b) shows the ratio between the scalar strahl temperature $T_s$ and the scalar core temperature $T_c$. This ratio slowly decreases with increasing $V_{sw}$ from $\sim 2.8$ in slow wind to $\sim 2$ at $V_{sw}\gtrsim 550\,\mathrm{km/s}$. Panel (c) shows the ratio between the scalar strahl temperature $T_s$ and the scalar halo temperature $T_h$. Like the other temperature ratios in Figure~\ref{fig_tratios}, $T_s/T_h$ shows only a small variation with $V_{sw}$. It assumes a value of $\sim 0.4$.

\begin{figure}[ht!]
\centering
\includegraphics[width=0.9\hsize]{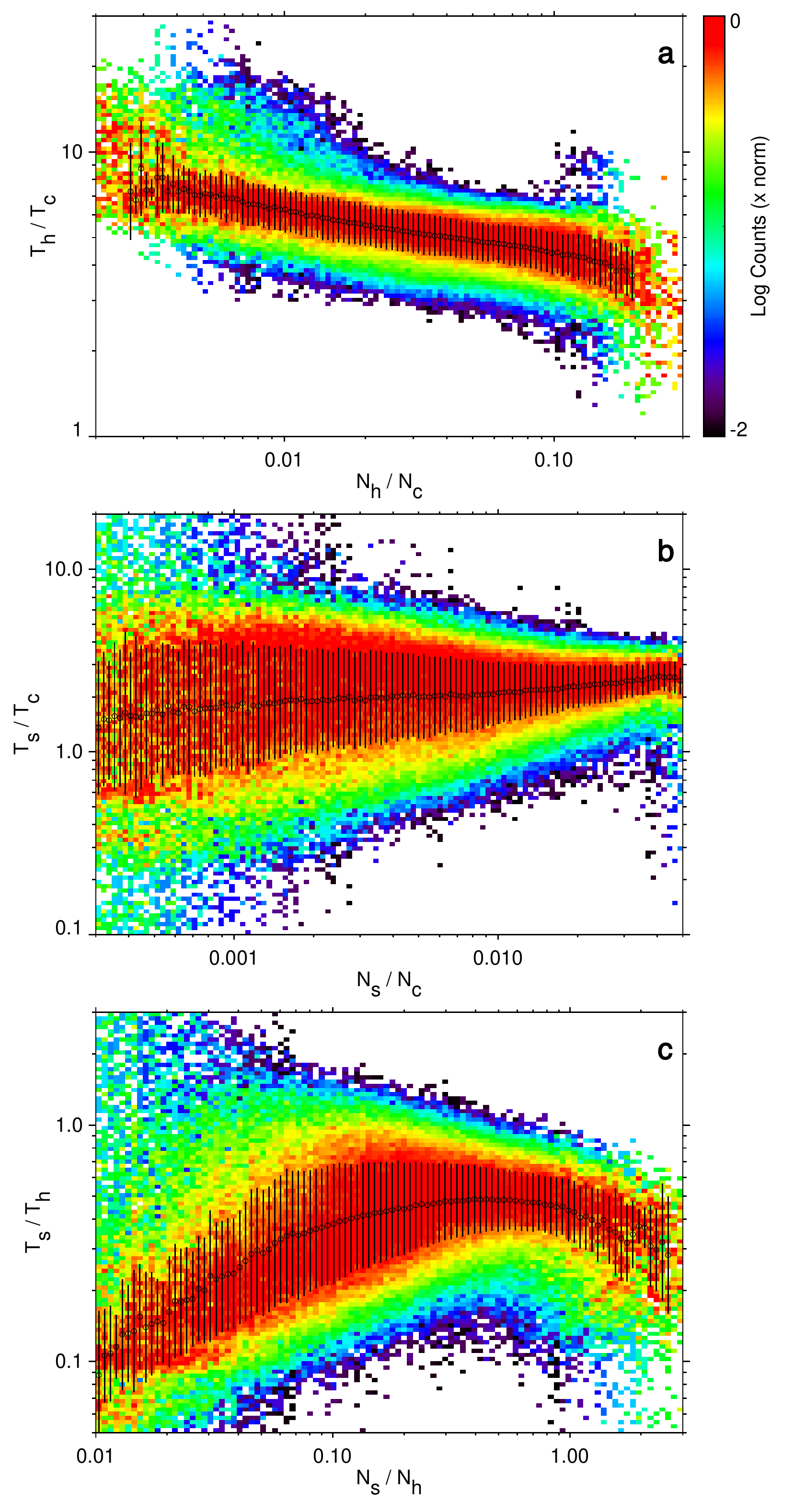}
\caption{Intercorrelations between electron temperature ratios and density ratios. (a) $T_h/T_s$ as a function of $N_h/N_c$, (b) $T_s/T_c$ as a function of $N_s/N_c$, and (c) $T_s/T_h$ as a function of $N_s/N_h$.}
\label{fig_chs_ratios}
\end{figure}

Figure~\ref{fig_chs_ratios} shows histograms of ratios between the scalar temperatures of the electron components as functions of their density ratios. These diagrams illustrate intercorrelations between the plotted quantities. Although we do not use $V_{sw}$ in Figure~\ref{fig_chs_ratios}, we still apply column-normalization and determine averages and standard deviations as in Figures~\ref{fig_totalparam1} through~\ref{fig_chs_ratios}. Panel (a) shows the correlation between $T_h/T_c$ and $N_h/N_c$. We see a clear trend from $T_h/T_c\sim 8$ at $N_h/N_c\sim 0.003$ to $T_h/T_c\sim 4$ at $N_h/N_c\sim 0.2$ with a small variability in each bin. Panel (b) shows the correlation between $T_s/T_c$ and $N_s/N_c$. These two ratios exhibit a less pronounced intercorrelation compared to the ratios shown in panel (a). We also note that the variability decreases with increasing $N_s/N_c$. The average $T_s/T_c$ slightly increases from $\sim 1.3$ at $N_s/N_c\sim 0.0003$ to $\sim 2.3$ at $N_s/N_c\sim 0.05$. Panel (c) displays the relationship between $T_s/T_h$ and $N_s/N_h$. The average ratio $T_s/T_h$ increases from $\sim 0.1$ at $N_s/N_h\sim 0.01$ to $\sim 0.5$ at $N_s/N_h\sim 0.4$, and then decreases again to $\sim 0.3$ at $N_s/N_h\sim2$.


\section{Discussion}

The total electron parameters in our dataset according to Figure~\ref{fig_totalparam1} are largely consistent with the known behavior in the solar wind. Fast wind exhibits significantly smaller electron densities (and thus, per quasi-neutrality, significantly smaller ion densities) than the slow solar wind. Our dataset also confirms that $T_e$ is generally smaller in fast-wind streams compared to slow wind \citep{Ogilvie:2000, Wilson:2018}. However, $T_e$ is highly variable in the slow solar wind, showing a general correlation with $V_{sw}$ in the slow solar wind up to $\sim$550\,km/s and an anti-correlation with $V_{sw}$ in the fast solar wind above 550\,km/s \citep{Maksimovic:2020}. It is interesting to note that the proton temperature shows an opposite and more pronounced trend (see Figure~\ref{fig_totalparam2}). This opposite behavior of $T_e$ and $T_p$ leads to a strong dependence of $T_e/T_p$ on $V_{sw}$. The ratio $T_e/T_p$ is of great interest for local plasma processes since, e.g., the damping rate of ion-acoustic waves and the energy partitioning of plasma heating are very sensitive to this parameter \citep{Howes:2006, Schekochihin:2009, Schekochihin:2019, Kawazura:2019, Kawazura:2020}. We also note that the density and temperature behavior cause a significant dependence of the average $\beta_e$ on $V_{sw}$, which is more pronounced than the dependence of $\beta_p$ on $V_{sw}$. This trend causes the average $\beta_e$ to cross the $\beta_e=1$ line between slow and fast wind streams.

In general, fast solar wind exhibits more non-equilibrium features in our dataset than slow solar wind. For instance, the total  eVDF as well as the core and (to a lesser extent) halo components exhibit more temperature anisotropy in fast wind than in slow wind (Figures~\ref{fig_totalparam1}, \ref{fig_corestats}, and \ref{fig_halostats}). Also the core drift is more pronounced in the fast solar wind than in slow solar wind. These results suggest that collisions, which occur more often in slow wind, likely play a role in the reduction of non-equilibrium kinetic features in the eVDF. In particular, collisions lead to a stronger effect in the core distribution than in the halo distribution.

While the average trends in the halo temperatures largely mirror the average trends in the core temperatures (Figure~\ref{fig_halostats}), the halo density is mostly constant, and the core-halo drift exhibits larger values in slow solar wind. The core-halo-strahl drift guarantees current balance (see Figure~\ref{fig_currentbalance}). The current contribution due to the Sunward core drift is compensated for by the anti-Sunward halo and strahl drifts. The halo is on average only slightly anisotropic with $T_{\perp}<T_{\parallel}$, which may be a consequence of the often suggested connection between strahl and halo electrons.

According to Figure~\ref{fig_strahlstats}, the strahl shows only small variations in $N_s$, $V_s\,\mathrm{sign(Q_{e\parallel})}$, and $T_{s\parallel}$ as functions of $V_{sw}$. However, $T_{s\perp}$ (and thus $T_s$ and $T_{\perp}/T_{\parallel}$ of the strahl) show a strong  anti-correlation with $V_{sw}$ on averge. These trends are consistent with the scenario of local scattering of strahl electrons towards larger $v_{\perp}$ \citep{Maksimovic:2005,Stverak:2009,Vasko:2019,Verscharen:2019}. In slow wind, the travel time of a plasma parcel from the Sun to 1\,au is longer than in fast wind, so that the scattering processes can act on the strahl electrons for a longer time as a possible explanation for the observed trend. This scattering scenario is also in agreement with Figure~\ref{fig_nratios} (b) and (to some extent) with Figure~\ref{fig_nratios} (c), which show that the relative strahl density is lower in the slow wind than in the fast wind, consistent with the scattering of strahl electrons into the halo. As a caveat to this interpretation, we note that the observed ratio of halo-to-core density ($N_h/N_c$) also increases with $V_{sw}$, which suggests that other processes in addition to  strahl scattering lead to a $V_{sw}$-independent trend in $N_h$ even though $N_c$ changes significantly with $V_{sw}$. In this context, we note that the temperature ratios between the electron components are mostly independent of $V_{sw}$ (see Figure~\ref{fig_tratios}), a constraint which must be fulfilled by kinetic models explaining the partitioning of particles and energies between core, halo, and strahl.

Figure~\ref{fig_chs_ratios} shows further interesting correlations between temperature ratios and density ratios of the electron components. Especially, panel (a) suggests that there is a clear correlation between $T_h/T_c$ and $N_h/N_c$ which requires theoretical explanation.


\section{Conclusions}

This paper presents a comprehensive analysis of the structure of the eVDF in the ambient solar wind at 1\,au, using data from the 3DP  and the WAVES experiment on board NASA's Wind spacecraft up to energies of 2-3\,keV.  

 The combination of data from the electrostatic analyzers measuring the actual eVDF and from the wave receivers measuring the quasi-thermal noise of the plasma lead to good estimates of Wind's spacecraft electrical potential \citep{Salem:2001, Pulupa:2014}, an unknown quantity otherwise. 

The correction for the spacecraft potential allows us to extract accurate properties of the eVDF in the solar wind. The properties of the three main electron populations in this energy range -- the core, halo and strahl -- are statistically characterized in great detail in the slow and fast solar wind.
The core and halo populations are modeled by the sum of a bi-Maxwellian distribution and a bi-$\kappa$ distribution, respectively.  The fit of the core-halo model to the observed distribution leads to a complete characterization of the core and halo populations. The strahl is extracted by substracting the core-halo model from the observed distributions, allowing us to integrate the strahl moments. In this way, we obtain a set of  moments and parameters for each population including temperature anisotropies and heat fluxes.

Our data-analysis algorithm is automated with the goal of analyzing several years worth of data statistically and building a database of accurate core, halo and strahl parameters in the solar wind. We here present statistics based on our initial four-year dataset (1995-1998).  This is to-date the best high-precision, large-scale electron dataset of the pristine solar wind. This dataset has already enabled further studies  \citep{Bale:2013, Pulupa:2014, Pulupa:2014b, Tong:2015, Chen:2016, He:2018, Yoon:2019, Verscharen:2019, Scudder:2021}. Our aim is to apply this technique to the entire 27 years of Wind data yielding the best electron dataset of almost two and half solar cycles in the pristine solar wind at 1\,au. However, this extension requires a careful re-evaluation of the microchannel-plate degradation over the course of the mission, which is beyond the scope of this work. In addition, the 27-year dataset will potentially include solar-cycle variations that introduce additional long-time variations to the statistical measures presented in this paper. Such a dataset will be a valuable reference for electron research in the context of ongoing electron measurements with the latest heliospheric space missions Parker Solar Probe and Solar Orbiter.


\begin{acknowledgements}
This work was supported by NASA grant NNX16AI59G and by NSF SHINE grant 1622498. D.V.~is supported by STFC Ernest Rutherford Fellowship ST/P003826/1 and STFC Consolidated Grant ST/S000240/1. 
The authors welcome collaborations, for which the dataset presented in this paper is available upon request. This work was discussed at the ESAC Solar Wind Electron Workshop in May 2019, which was supported by the Faculty of the European Space Astronomy Centre (ESAC). 
\end{acknowledgements}

%
%





\begin{appendix}
\section{Statistics} \label{appendix}

In this appendix, we provide numerical and tabulated statistical properties (averages and standard deviations) of the electron parameters presented in Figures \ref{fig_totalparam1}, \ref{fig_totalparam2}, \ref{fig_corestats}, \ref{fig_halostats}, \ref{fig_strahlstats}, \ref{fig_nratios}, and \ref{fig_tratios}.

\begin{figure}[ht]
\centering
\includegraphics[width=0.9\hsize]{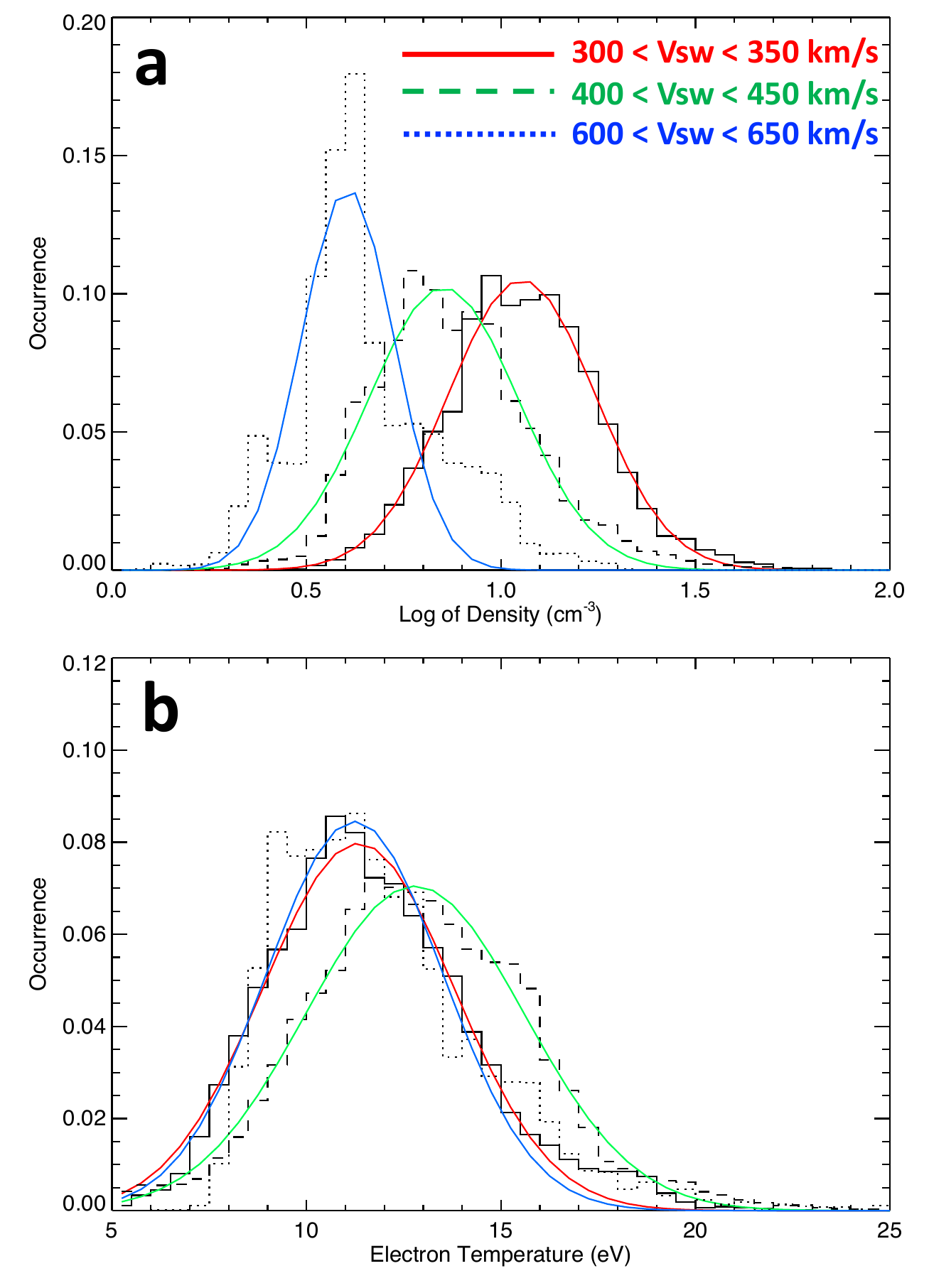}
\caption{Histograms of (a) the logarithm of the electron density and (b) the electron temperature for three different bins of solar wind speed $V_{sw}$. The colored lines represents Gaussian fits to the observed distributions.}
\label{histo_stats_Ne_Te}
\end{figure}

For the numbers presented in the tables in this appendix, we choose a binsize in $V_{sw}$ of 50\,km/s, rather that the 5\,km/s used in the 2D histograms in Section 4. For each variable $x$, $\langle x \rangle$ is the Gaussian peak centroid and $\sigma_x$ the half width. These statistics $\langle x \rangle \pm \sigma_x$ pertain to the characterization of the mode (most frequently occurring) locale using a Gaussian fit to the distribution of a variable $x$ in bins of solar wind speed $V_{sw}$, as it is standard when characterizing fluctuating variables. This characterization is more appropriate here than regular arithmetic means and standard deviations.

Electron densities $N_i$ are in cm$^{-3}$, drift velocities $V_{di}$ in km/s, temperatures $T_i$ in eV, and heat flux $Q_i$ in $\mu$W/m$^{-2}$; the indices $i = e, \;c, \;h, \;\rm{or}\; s$ represent total electron, core, halo or strahl parameters respectively. If $x = \log_{10} y$, then the mode of $y$ is characterized by $\langle y \rangle = 10^{\langle x \rangle}$ and error bars $\sigma_y = 10^{[\langle x \rangle \pm \sigma_x]}$.

An example of the Gaussian fit method is illustrated in Figure \ref{histo_stats_Ne_Te}, which displays histograms of the logarithm of the electron density $\log\,N_e$ (panel a) and histograms of the electron temperature $T_e$ (panel b) for three different bins in $V_{sw}$.  The solid line represents data for $V_{sw} \in [300, 350]$\,km/s, the dashed line represents data for $V_{sw} \in [400, 450]$\,km/s and the dotted line represents data for $V_{sw} \in [600, 650]$\,km/s.  We fit the three histograms with a Gaussian distribution. This is done for each of the parameters $x$ shown in Figs.~\ref{fig_totalparam1}, \ref{fig_totalparam2}, \ref{fig_corestats}, \ref{fig_halostats}, \ref{fig_strahlstats}, \ref{fig_nratios}, and \ref{fig_tratios}. We report the Gaussian peak centroid $\langle x \rangle$ and half width $\sigma_x$  in the tables \ref{table_fig08} to \ref{table_fig15} below.

\begin{table*}
\caption{\small Statistics of parameters presented in Figure \ref{fig_totalparam1}.}
\label{table_fig08}
\centering
\begin{tabular}{|l|r|r|r|r|r|r|}
\hline
$V_{sw}$  &  $\log_{10} N_e$  &  $T_{e\parallel}$  &  $T_{e\perp}$  &  $T_e$  &  $T_{e\perp}/T_{e\parallel}$  &  $\log_{10} Q_e$ \\
  &  &  &  &  &  & \\
 \hline
  &  &  &  &  &  & \\
250 -- 300  &  $1.10 \pm 0.15$  &  $9.92 \pm 1.39$   &  $9.71 \pm 1.22$  &  $9.81 \pm 1.30$  &  $0.99 \pm 0.04$  &  $0.59 \pm 0.21$ \\
300 -- 350  &  $1.05 \pm 0.19$  &  $11.83 \pm 2.55$  &  $11.39 \pm 2.33$  &  $11.56 \pm 2.40$  & $0.97 \pm 0.05$   & $0.66 \pm 0.32$  \\
350 -- 400  &  $0.96 \pm 0.18$  &  $12.56 \pm 2.98$  &  $11.83 \pm 2.74$  &  $12.10 \pm 2.83$  &  $0.95 \pm 0.06$  &  $0.71 \pm 0.29$   \\
400 -- 450  &  $0.87 \pm 0.18$  &  $13.52 \pm 2.91$  &  $12.40 \pm 2.61$  &  $12.83 \pm 2.75$  &  $0.92 \pm 0.07$  &  $0.78 \pm 0.27$  \\
450 -- 500  &  $0.81 \pm 0.19$  &  $14.24 \pm 2.94$  &  $12.60 \pm 2.69$  &  $13.18 \pm 2.74$  &  $0.89 \pm 0.08$  &  $0.77 \pm 0.30$   \\
500 -- 550  &  $0.73 \pm 0.19$  &  $14.46 \pm 2.98$  &  $12.29 \pm 2.55$  &  $13.04 \pm 2.61$  &  $0.86 \pm 0.08$  &  $0.74 \pm 0.24$   \\
550 -- 600  &  $0.67 \pm 0.17$  &  $15.48 \pm 3.04$  &  $12.58 \pm 2.37$  &  $13.49 \pm 2.35$  &  $0.83 \pm 0.09$  &   $0.74 \pm 0.22$ \\
600 -- 650  &  $0.66 \pm 0.18$  &  $13.72 \pm 2.27$  &  $10.69 \pm 1.82$  &  $11.67 \pm 1.87$  &  $0.79 \pm 0.10$  &   $0.71 \pm 0.21$  \\
650 -- 700  &  $0.62 \pm 0.13$  &  $12.68 \pm 1.80$  &  $8.90 \pm1.10$  &  $10.05 \pm1.29$  &  $0.74 \pm 0.08$  &   $0.65 \pm 0.21$ \\
700 -- 750  &  $0.56 \pm 0.11$  &  $12.37 \pm 1.81$  &  $8.48 \pm 0.83$  &  $9.55 \pm 0.90$  &  $0.72 \pm 0.08$  &   $0.55 \pm 0.18$ \\
750 -- 800  &  $0.57 \pm 0.08$ &  $10.86 \pm 0.56$  &  $8.51 \pm 0.78$  &  $9.64 \pm 0.90$  &  $0.70 \pm 0.09$  &  $0.45 \pm 0.15$  \\
  &  &  &  &  &  & \\
\hline
\end{tabular}
\end{table*}

\begin{table*}
\caption{\small Statistics of parameters presented in Figure \ref{fig_totalparam2}.}
\label{table_fig09}
\centering
\begin{tabular}{|l|r|r|r|r|}
\hline
$V_{sw}$  &  $T_p$  &  $\log_{10} T_e/T_p$  &  $\log_{10} \beta_p$  &  $\log_{10} \beta_e$  \\
  &  &  &  &  \\
 \hline
  &  &  &  &   \\
250 -- 300  &  $1.84 \pm 0.67$  &  $0.72 \pm 0.15$ &  $-0.30 \pm 0.30$ &  $0.49 \pm 0.34$  \\
300 -- 350  &  $2.93 \pm 1.23$  &  $0.64 \pm 0.23$ &  $-0.33 \pm 0.30$ &  $0.33 \pm 0.31$  \\
350 -- 400  &  $4.60 \pm 2.14$  &  $0.42 \pm 0.23$ &  $-0.28 \pm 0.32$ &  $0.15 \pm 0.31$  \\
400 -- 450  &  $6.84 \pm 2.62$  &  $0.27 \pm 0.22$ &  $-0.24 \pm 0.30$ &  $0.04 \pm 0.27$  \\
450 -- 500  &  $9.36 \pm 4.19$  &  $0.11 \pm 0.21$ &  $-0.19 \pm 0.30$ &  $-0.05 \pm 0.29$ \\
500 -- 550  &  $12.31 \pm 4.50$  &  $ 0.00 \pm 0.19$ &  $-0.14 \pm 0.27$ &  $-0.11 \pm 0.26$  \\
550 -- 600  &  $14.93 \pm 4.53$  &  $-0.09 \pm 0.19$ &  $-0.09 \pm 0.23$ &  $-0.13 \pm 0.22$  \\
600 -- 650  &  $17.18 \pm 4.53$  &  $-0.23 \pm 0.16$ &  $-0.07 \pm 0.22$ &  $-0.23 \pm 0.20$  \\
650 -- 700  &  $20.20 \pm 4.36$  &  $-0.32 \pm 0.12$ &  $-0.07 \pm 0.20$ &  $-0.34 \pm 0.16$  \\
700 -- 750  &  $22.72 \pm 4.91$  &  $-0.37 \pm 0.12$ &  $-0.07 \pm 0.19$ &  $-0.39 \pm 0.12$  \\
750 -- 800  &  $25.57 \pm 2.93$  &  $-0.44 \pm 0.08$ &  $0.01 \pm 0.17$ &  $-0.39 \pm 0.10$  \\
  &  &  &  &   \\
\hline
\end{tabular}
\end{table*}

\begin{table*}
\caption{\small Statistics of core parameters presented in Figure \ref{fig_corestats}.}
\label{table_fig10}
\centering
\begin{tabular}{|l|r|r|r|r|r|r|}
\hline
$V_{sw}$  &  $\log_{10} N_c$  &  $V_{dc}\,\rm{sign}[Q_e]$  &  $T_{c\parallel}$  &  $T_{c\perp}$ &  $T_c$  &  $T_{c\perp}/T_{c\parallel}$ \\
  &  &  &  &  &  & \\
 \hline
  &  &  &  &  &  & \\
250 -- 300  &  $1.10 \pm 0.16$  &  $-25.17 \pm 34.75$  &  $8.76 \pm 1.14$  &  $8.87 \pm 1.07$  &  $8.84 \pm 1.06$  &  $1.02 \pm 0.04$  \\
300 -- 350  &  $1.04 \pm 0.19$  &  $-42.33 \pm 42.28$  &  $9.92 \pm 1.87$  &  $10.08 \pm 1.94$  &  $10.02 \pm 1.89$  &  $1.02 \pm 0.05$  \\
350 -- 400  &  $0.95 \pm 0.19$  &  $-53.77 \pm 45.85$  &  $9.96 \pm 1.89$  &  $10.10 \pm 2.01$  &  $10.07 \pm 1.96$  &  $1.01 \pm 0.05$  \\
400 -- 450  &  $0.85 \pm 0.19$  &  $-67.60 \pm 48.31$  &  $10.22 \pm 1.84$  &  $10.33 \pm 1.87$  &  $10.32 \pm 1.89$  &  $1.01 \pm 0.05$  \\
450 -- 500  &  $0.79 \pm 0.20$  &  $-77.05 \pm 55.41$  &  $10.36 \pm 1.80$  &  $10.17 \pm 1.75$  &  $10.25 \pm 1.75$  &  $0.98 \pm 0.06$  \\
500 -- 550  &  $0.70 \pm 0.20$  &  $-83.09 \pm 52.81$  &  $10.11 \pm 1.85$  &  $9.73 \pm 1.80$  &  $9.82 \pm 1.74$  &  $0.96 \pm 0.07$  \\
550 -- 600  &  $0.63 \pm 0.18$  &  $-91.60 \pm 56.19$  &  $10.37 \pm 1.90$  &  $9.40 \pm 1.55$  &  $9.64 \pm 1.57$  &  $0.93 \pm 0.08$  \\
600 -- 650  &  $0.62 \pm 0.18$  &  $-93.16 \pm 56.20$  &  $8.67 \pm 1.29$  &  $8.15 \pm 1.23$  &  $8.39 \pm 1.25$  &  $0.91 \pm 0.08$  \\
650 -- 700  &  $0.59 \pm 0.14$  &  $-100.19 \pm 57.67$  &  $7.83 \pm 1.03$  &  $6.74 \pm 0.78$  &  $7.12 \pm 0.87$  &  $0.87 \pm 0.07$  \\
700 -- 750  &  $0.53 \pm 0.11$  &  $-101.72 \pm 56.31$  &  $7.84 \pm 1.14$  &  $6.60 \pm 0.79$  &  $7.01 \pm 0.84$  &  $0.84 \pm 0.08$  \\
750 -- 800  &  $0.53 \pm 0.08$  &  $-110.66 \pm 50.58$  &  $7.22 \pm 0.60$  &  $6.62 \pm 0.85$  &  $6.95 \pm 0.91$  &  $0.87 \pm 0.04$  \\
  &  &  &  &  &  & \\
\hline
\end{tabular}
\end{table*}

\begin{table*}
\caption{\small Statistics of halo parameters presented in Figure \ref{fig_halostats}.}
\label{table_fig11}
\centering
\begin{tabular}{|l|r|r|r|r|r|r|}
\hline
$V_{sw}$  &  $\log_{10} N_h$  &  $V_{dh}\,\rm{sign}[Q_e]$  &  $T_{h\parallel}$  &  $T_{h\perp}$ &  $T_h$  &  $T_{h\perp}/T_{h\parallel}$ \\
  &  &  &  &  &  & \\
 \hline
  &  &  &  &  &  & \\
250 -- 300  &  $-0.70 \pm 0.21$  &  $307.65 \pm 543.90$  &  $53.95 \pm 8.62$  &  $52.33 \pm 6.59$  &  $52.59 \pm 7.07$  &  $0.96 \pm 0.08$  \\
300 -- 350  &  $-0.56 \pm 0.23$  &  $333.63 \pm 510.00$  &  $53.66 \pm 9.86$  &  $50.82 \pm 9.29$  &  $51.81 \pm 9.19$  &  $0.95 \pm 0.07$  \\
350 -- 400  &  $-0.55 \pm 0.24$  &  $355.47 \pm 496.91$  &  $53.88 \pm 9.77$  &  $50.74 \pm 9.45$  &  $51.81 \pm 9.21$  &  $0.94 \pm 0.07$  \\
400 -- 450  &  $-0.53 \pm 0.23$  &  $394.39 \pm 495.21$  &  $54.84 \pm 9.94$  &  $51.37 \pm 9.04$  &  $52.59 \pm 9.06$  &  $0.94 \pm 0.08$  \\
450 -- 500  &  $-0.54 \pm 0.22$  &  $378.25 \pm 512.53$  &  $55.71 \pm 10.39$  &  $52.16 \pm 9.14$  &  $53.46 \pm 9.31$  &  $0.93 \pm 0.08$  \\
500 -- 550  &  $-0.55 \pm 0.24$  &  $340.70 \pm 499.48$  &  $52.75 \pm 11.06$  &  $48.88 \pm 9.86$  &  $50.21 \pm 9.99$  &  $0.93 \pm 0.08$  \\
550 -- 600  &  $-0.56 \pm 0.23$  &  $325.37 \pm 538.74$  &  $50.96 \pm 11.98$  &  $47.31 \pm 10.06$  &  $48.62 \pm 10.50$  &  $0.92 \pm 0.08$  \\
600 -- 650  &  $-0.59 \pm 0.26$  &  $203.03 \pm 565.90$  &  $44.41 \pm 9.75$  &  $40.34 \pm 8.23$  &  $41.81 \pm 8.73$  &  $0.91 \pm 0.08$  \\
650 -- 700  &  $-0.62 \pm 0.23$  &  $40.40 \pm 542.46$  &  $39.52 \pm 10.21$  &  $35.58 \pm 9.07$  &  $37.02 \pm 9.80$  &  $0.90 \pm 0.08$  \\
700 -- 750  &  $-0.67 \pm 0.23$  &  $-160.88 \pm 532.99$  &  $37.09 \pm 9.92$  &  $33.57 \pm 8.57$  &  $34.86 \pm 9.26$  &  $0.90 \pm 0.08$  \\
750 -- 800  &  $-0.66 \pm 0.22$  &  $-341.36 \pm 418.05$  &  $32.06 \pm 7.39$  &  $30.34 \pm 6.24$  &  $30.30 \pm 7.64$  &  $0.93 \pm 0.09$  \\
  &  &  &  &  &  & \\
\hline
\end{tabular}
\end{table*}


\begin{table*}
\caption{\small Statistics of strahl parameters presented in Figure \ref{fig_strahlstats}.}
\label{table_fig12}
\centering
\begin{tabular}{|l|r|r|r|r|r|r|}
\hline
$V_{sw}$  &  $\log_{10} N_s$  &  $V_{ds}\,\rm{sign}[Q_e]$  &  $T_{s\parallel}$  &  $T_{s\perp}$ &  $T_s$  &  $T_{s\perp}/T_{s\parallel}$ \\
  &  &  &  &  &  & \\
 \hline
  &  &  &  &  &  & \\
250 -- 300  &  $-1.45 \pm 0.34$  &  $4466.05 \pm 541.09$  &  $12.85 \pm 3.76$  &  $29.55 \pm 13.16$  &  $22.63 \pm 9.52$  &  $2.87 \pm 0.60$  \\
300 -- 350  &  $-1.38 \pm 0.36$  &  $4655.05 \pm 598.71$  &  $12.81 \pm 3.24$  &  $30.31 \pm 20.22$  &  $23.44 \pm 11.23$  &  $2.98 \pm 0.78$  \\
350 -- 400  &  $-1.33 \pm 0.34$  &  $4618.31 \pm 566.12$  &  $12.84 \pm 3.26$  &  $28.42 \pm 21.21$  &  $22.22 \pm 11.32$  &  $2.76 \pm 0.78$  \\
400 -- 450  &  $-1.26 \pm 0.32$  &  $4660.95 \pm 579.84$  &  $13.06 \pm 3.29$  &  $26.86 \pm 15.72$  &  $21.32 \pm 10.33$  &  $2.59 \pm 0.82$  \\
450 -- 500  &  $-1.27 \pm 0.33$  &  $4790.17 \pm 637.31$  &  $13.47 \pm 3.59$  &  $24.31 \pm 13.98$  &  $20.38 \pm 10.06$  &  $2.27 \pm 0.75$  \\
500 -- 550  &  $-1.25 \pm 0.30$  &  $4760.16 \pm 630.74$  &  $13.51 \pm 3.01$  &  $22.60 \pm 14.32$  &  $19.56 \pm 9.17$  &  $2.09 \pm 0.69$  \\
550 -- 600  &  $-1.25 \pm 0.28$  &  $4748.81 \pm 617.19$  &  $13.43 \pm 2.73$  &  $19.56 \pm 13.48$  &  $18.03 \pm 8.57$  &  $1.88 \pm 0.66$  \\
600 -- 650  &  $-1.20 \pm 0.23$  &  $4677.90 \pm 604.30$  &  $13.66 \pm 2.29$  &  $18.47 \pm 10.27$  &  $17.45 \pm 7.14$  &  $1.59 \pm 0.54$  \\
650 -- 700  &  $-1.18 \pm 0.19$  &  $4628.23 \pm 568.30$  &  $13.98 \pm 2.11$  &  $16.34 \pm 8.97$  &  $16.37 \pm 6.25$  &  $1.34 \pm 0.46$  \\
700 -- 750  &  $-1.19 \pm 0.18$  &  $4629.94 \pm 560.60$  &  $13.51 \pm 2.09$  &  $15.93 \pm 7.53$  &  $15.31 \pm 5.25$  &  $1.29 \pm 0.44$  \\
750 -- 800  &  $-1.23 \pm 0.16$  &  $4565.21 \pm 471.56$  &  $12.52 \pm 1.20$  &  $13.80 \pm 6.85$  &  $14.35 \pm 4.20$  &  $1.27 \pm 0.39$  \\
  &  &  &  &  &  & \\
\hline
\end{tabular}
\end{table*}

\begin{table*}
\caption{\small Statistics of density ratios presented in Figure \ref{fig_nratios}.}
\label{table_fig14}
\centering
\begin{tabular}{|l|r|r|r|}
\hline
$V_{sw}$  &  $\log_{10} N_h/N_c$  &  $\log_{10} N_s/N_c$  &  $\log_{10} N_s/N_h$   \\
  &  &  &   \\
 \hline
  &  &  &   \\
250 -- 300  &  $-1.82 \pm 0.25$  &  $-2.57 \pm 0.43$  &  $-0.75 \pm 0.40$   \\
300 -- 350  &  $-1.62 \pm 0.28$  &  $-2.46 \pm 0.46$  &  $-0.80 \pm 0.40$   \\
350 -- 400  &  $-1.51 \pm 0.29$  &  $-2.30 \pm 0.41$  &  $-0.75 \pm 0.39$   \\
400 -- 450  &  $-1.38 \pm 0.24$  &  $-2.11 \pm 0.34$  &  $-0.71 \pm 0.39$   \\
450 -- 500  &  $-1.34 \pm 0.23$  &  $-2.04 \pm 0.35$  &  $-0.68 \pm 0.43$  \\
500 -- 550  &  $-1.25 \pm 0.21$  &  $-1.94 \pm 0.31$  &  $-0.67 \pm 0.43$  \\
550 -- 600  &  $-1.20 \pm 0.21$  &  $-1.88 \pm 0.30$  &  $-0.67 \pm 0.42$   \\
600 -- 650  &  $-1.22 \pm 0.23$  &  $-1.80 \pm 0.26$  &  $-0.59 \pm 0.41$  \\
650 -- 700  &  $-1.21 \pm 0.23$  &  $-1.79 \pm 0.25$  &  $-0.56 \pm 0.34$  \\
700 -- 750  &  $-1.21 \pm 0.22$  &  $-1.74 \pm 0.22$  &  $-0.52 \pm 0.31$  \\
750 -- 800  &  $-1.16 \pm 0.19$  &  $-1.75 \pm 0.21$  &  $-0.53 \pm 0.27$ \\
  &  &  &   \\
\hline
\end{tabular}
\end{table*}

\begin{table*}
\caption{\small Statistics of temperature ratios presented in Figure \ref{fig_tratios}.}
\label{table_fig15}
\centering
\begin{tabular}{|l|r|r|r|}
\hline
$V_{sw}$  &  $\log_{10} T_h/T_c$  &  $\log_{10} T_s/T_c$  &  $\log_{10} T_s/T_h$   \\
  &  &  &   \\
 \hline
  &  &  &   \\
250 -- 300  &  $5.90 \pm 0.92$  &  $2.58 \pm 1.11$  &  $0.45 \pm 0.18$  \\
300 -- 350  &  $5.23 \pm 0.97$  &  $2.42 \pm 0.98$  &  $0.48 \pm 0.19$  \\
350 -- 400  &  $5.20 \pm 0.94$  &  $2.31 \pm 0.97$  &  $0.45 \pm 0.19$  \\
400 -- 450  &  $5.11 \pm 0.93$  &  $2.14 \pm 0.89$  &  $0.43 \pm 0.18$  \\
450 -- 500  &  $5.12 \pm 0.83$  &  $2.04 \pm 0.85$  &  $0.40 \pm 0.17$  \\
500 -- 550  &  $5.05 \pm 0.99$  &  $2.15 \pm 0.91$  &  $0.42 \pm 0.18$ \\
550 -- 600  &  $5.04 \pm 1.00$  &  $1.99 \pm 0.90$  &  $0.39 \pm 0.19$  \\
600 -- 650  &  $5.04 \pm 1.03$  &  $2.19 \pm 0.96$  &  $0.43 \pm 0.20$  \\
650 -- 700  &  $4.82 \pm 1.01$  &  $2.23 \pm 0.88$  &  $0.46 \pm 0.18$ \\
700 -- 750  &  $4.77 \pm 1.08$  &  $2.18 \pm 0.77$  &  $0.45 \pm 0.17$ \\
750 -- 800  &  $4.53 \pm 0.92$  &  $2.10 \pm 0.70$  &  $0.43 \pm 0.16$ \\
  &  &  &   \\
\hline
\end{tabular}
\end{table*}


\end{appendix}

\end{document}